\documentclass[12pt,eqno,epsf]{article}
\usepackage{amsmath,amssymb,graphicx,slashbox}
\numberwithin{equation}{section}

\def\mydate{November 28, 2011}
\def\ignore#1{{}}

\newcounter{sxn}

\newcounter{axn}

\date{}

\newdimen\mybaselineskip
\renewcommand{\thefootnote}{\arabic{footnote}}

\newcommand{\beeq}{\begin{equation}}
\newcommand{\eneq}{\end{equation}}
\newcommand{\beqn}{\begin{eqnarray}}
\newcommand{\eeqn}{\end{eqnarray}}

\newcommand{\alp}{\alpha}
\newcommand{\bt}{\beta}
\newcommand{\gm}{\gamma}

\newcommand{\dlt}{\delta}

\newcommand{\ep}{\epsilon}
\newcommand{\tht}{\theta}

\newcommand{\kp}{\kappa}
\newcommand{\lmd}{\lambda}
\newcommand{\Lmd}{\Lambda}
\newcommand{\sgm}{\sigma}
\newcommand{\Sgm}{\Sigma}
\newcommand{\vph}{\varphi}
\newcommand{\omg}{\omega}
\newcommand{\Omg}{\Omega}
\newcommand{\dalp}{\dot{\alpha}}
\newcommand{\dbt}{\dot{\beta}}
\newcommand{\dgm}{\dot{\gamma}}

\newcommand{\be}{\begin{equation}}
\newcommand{\ee}{\end{equation}}
\newcommand{\bea}{\begin{eqnarray}}
\newcommand{\eea}{\end{eqnarray}}
\newcommand{\eql}{\!\!\!&=\!\!\!&}

\newcommand{\defa}{\!\!\!&\equiv\!\!\!&}

\newcommand{\lrder}{\stackrel{\leftrightarrow}{\partial}}

\newcommand{\tl}[1]{\tilde{#1}}
\newcommand{\bdm}[1]{{\mbox{\boldmath $#1$}}}

\newcommand{\der}{\partial}
\newcommand{\dr}{\!\!d}
\newcommand{\hc}{{\rm h.c.}}
\newcommand{\ie}{{\it i.e.}}

\newcommand{\dsc}{\dlt_{\rm sc}}

\newcommand{\vev}[1]{\langle #1 \rangle}

\newcommand{\brkt}[1]{\left( #1 \right)}
\newcommand{\brc}[1]{\left\{ #1 \right\}}
\newcommand{\sbk}[1]{\left[ #1 \right]}

\renewcommand{\Re}{{\rm Re}\,}
\renewcommand{\Im}{{\rm Im}\,}

\newcommand{\cD}{{\cal D}}
\newcommand{\cE}{{\cal E}}

\newcommand{\cH}{{\cal H}}

\newcommand{\cL}{{\cal L}}

\newcommand{\cO}{{\cal O}}

\newcommand{\cV}{{\cal V}}
\newcommand{\cW}{{\cal W}}
\newcommand{\cX}{{\cal X}}
\newcommand{\cY}{{\cal Y}}

\newcommand{\tpsi}{\tl{\psi}}

\begin{document}
\thispagestyle{empty}

\baselineskip=12pt

{\small \noindent \mydate    
\hfill }

{\small \noindent \hfill  KEK-TH-1475}

\baselineskip=35pt plus 1pt minus 1pt

\vskip 1.5cm

\begin{center}
{\LARGE\bf Direct relation of linearized supergravity} \\
{\LARGE\bf to superconformal formulation}\\ 
%{\LARGE\bf Complete identification of fields} \\
%{\LARGE\bf in linearized supergravity with those} \\ 
%{\LARGE\bf in superconformal formulation}\\

\vspace{1.5cm}
\baselineskip=20pt plus 1pt minus 1pt

\normalsize

{\large\bf Yutaka\ Sakamura}$\!${\def\thefootnote{\fnsymbol{footnote}}
\footnote[1]{\tt e-mail address: sakamura@post.kek.jp}}

\vspace{.3cm}
{\small \it KEK Theory Center, Institute of Particle and Nuclear Studies, 
KEK, \\ Tsukuba, Ibaraki 305-0801, Japan} \\ \vspace{3mm}
{\small \it Department of Particles and Nuclear Physics, \\
The Graduate University for Advanced Studies (Sokendai), \\
Tsukuba, Ibaraki 305-0801, Japan} 
\end{center}

\vskip 1.0cm
\baselineskip=20pt plus 1pt minus 1pt

\begin{abstract}
We modify the four-dimensional $N=1$ linearized supergravity in a way that 
components in each superfield are completely identified with 
fields in the full superconformal formulation. 
This identification makes it possible to use both formulations of supergravity 
in a complementary manner. 
It also provides a basis for an extension 
to higher-dimensional supergravities. 
\end{abstract}

%%%%%%%%%%%%%
%\centerline{PACS: }
%%%%%%%%%%%%%

\newpage

\section{Introduction}
The superconformal formulation of supergravity (SUGRA) is 
a powerful and systematic method 
for constructing various SUGRA actions~\cite{Kaku:1978nz,Kaku:1978ea,
Ferrara:1978rk,Kugo:1982cu}. 
Most of the known off-shell SUGRA actions are reproduced 
by this formulation. 
It has also been extended to the five-dimensional (5D) 
case~\cite{5D_Kugo,Kugo:2002js}, 
which is useful to discuss the brane-world scenario 
based on general 5D SUGRA. 
Although the actions are obtained in a systematic way, 
their explicit expressions are lengthy and awkward due to a number of
auxiliary fields. 
Especially the couplings between the matter and the SUGRA fields 
(\ie, the vierbein, the gravitino, etc.) are complicated. 

Linearized supergravity~\cite{Ferrara:1977mv,Siegel:1978mj} is easier to deal with 
because it is described in terms of superfields on the ordinary superspace. 
%The actions are expressed in terms of the superfields 
%in compact forms. 
%In fact, the off-shell supergravity actions were first constructed 
%at linearized order, and extended to the nonlinear order. 
%Although we cannot use this formalism beyond the linearized order 
%for the SUGRA fields, 
It is powerful for some calculations 
because the ordinary superfield techniques are applicable 
just as in the global supersymmetry (SUSY) case.
An extension to 5D case for the minimal set-up 
was done in Ref.~\cite{Linch:2002wg}, 
and it makes it possible to calculate the SUGRA loop contributions 
in the 5D brane-world models~\cite{Buchbinder:2003qu},
%\footnote{
%For some quantities, the SUGRA loop contributions are finite 
%due to the five-dimensional locality.}
keeping the $N=1$ SUSY off-shell structure. 
%This 5D extension is based on the 4D formulation. 
On the other hand, we cannot use this formalism for calculations 
beyond the linearized order in the SUGRA fields. 
The full SUGRA formulation, such as the superconformal formulation, 
is necessary for them. 

Therefore it will be useful to combine the two formulations 
in a complementary manner. 
In fact, it is pointed out in Ref.~\cite{Ferrara:1977mv,Siegel:1978mj} that 
the linearized SUGRA transformations contain some of the superconformal 
transformations at the linearized level. 
Although both formulations are self-consistent, 
an explicit relation between them has not been provided so far. 
This is the main obstacle to the complementary use of them. 

In this paper, we will modify the linearized SUGRA formulation 
and provide a complete identification of component fields 
in each superfield with fields in the superconformal formulation 
developed in Ref.~\cite{Kugo:1982cu}. 
This identification also provides a basis for 
an extension to higher-dimensional SUGRA. 

The paper is organized as follows.
In Sec.~\ref{sf_trf}, we consider superfield transformations 
which are identified with 
the linearized superconformal transformations. 
In Sec.~\ref{comp:identify}, we translate such transformation laws 
into those for component fields, and identify the fields and 
the transformation parameters with those in the superconformal formulation
of Ref.~\cite{Kugo:1982cu}. 
In Sec.~\ref{Action_fml}, we construct the invariant action formulae 
in terms of the superfields, which are consistent with those 
in Ref.~\cite{Kugo:1982cu}. 
Sec.~\ref{summary} is devoted to the summary. 
In Appendix~\ref{comp_expr}, we provide explicit component expressions of 
some superfields in the text, 
and in Appendix~\ref{FD_action_fml}, we collect the invariant action formulae 
in Ref.~\cite{Kugo:1982cu} in our notations.

\section{Superfield transformations} \label{sf_trf}
In this section, we consider the transformation laws of $N=1$ superfields. 
We assume that the background geometry is a flat 4D Minkowski spacetime. 
Basically we use the spinor notations of Ref.~\cite{Wess:1992cp}, 
except for the metric and the spinor derivatives. 
We take the background metric as $\eta_{\mu\nu}=(1,-1,-1,-1)$ 
so as to match it to that of Ref.~\cite{Kugo:1982cu}, and 
we define the spinor derivatives~$D_\alp$ and $\bar{D}_{\dalp}$ as 
\be
 D_\alp \equiv \frac{\der}{\der\tht^\alp}-i\brkt{\sgm^\mu\bar{\tht}}_\alp\der_\mu, 
 \;\;\;\;\;
 \bar{D}_{\dalp} \equiv -\frac{\der}{\der\bar{\tht}^{\dalp}}
 +i\brkt{\tht\sgm^\mu}_{\dalp}\der_\mu,  \label{def:DbD}
\ee
which satisfy $\brc{D_\alp,\bar{D}_{\dalp}}
=2i\sgm^\mu_{\alp\dalp}\der_\mu$.\footnote{ 
In the notation of Ref.~\cite{Wess:1992cp}, the spinor derivatives satisfy 
$\brc{D_\alp,\bar{D}_{\dalp}}=-2i\sgm^\mu_{\alp\dalp}\der_\mu$. 
Then, the (global) SUSY generators~$Q_\alp$ 
and $\bar{Q}_{\dalp}$, which anticommute with $D_\alp$ and $\bar{D}_{\dalp}$, 
satisfy $\brc{Q_\alp,\bar{Q}_{\dalp}}=2i\sgm^\mu_{\alp\dalp}\der_\mu$. 
This leads to the SUSY algebra with an opposite sign to the usual one. 
(See Chapter IV of Ref.~\cite{Wess:1992cp}.) }

\subsection{Super-diffeomorphism}
We begin with a brief review of the formulation developed 
in Ref.~\cite{Siegel:1978mj}. 
(A compact review is also provided in Ref.~\cite{Linch:2002wg}.)
First we consider the super-diffeomorphism transformation acting on 
a chiral superfield~$\Phi$. 
It is expressed as 
\be
 \dlt\Phi = \Lmd^\alp D_\alp\Phi+\Lmd^\mu\der_\mu\Phi, 
\ee
where $\Lmd^\alp$ and $\Lmd^\mu$ are a spinor and a vector superfields, 
respectively. 
We require that 
this transformation~$\dlt$ preserves the chiral condition, 
\be
 \bar{D}_{\dalp}\Phi = 0.  \label{cond:chiral}
\ee
Then we obtain 
\be
 \bar{D}_{\dalp}\dlt\Phi = \bar{D}_{\dalp}\Lmd^\alp D_\alp\Phi
 +\brkt{-2i\Lmd^\alp\sgm_{\alp\dalp}^\mu+\bar{D}_{\dalp}\Lmd^\mu}\der_\mu\Phi
 = 0. 
\ee
We have used (\ref{cond:chiral}). 
Thus we find that
\be
 \bar{D}_{\dalp}\Lmd^\alp = 0, \;\;\;\;\;
 -2i\Lmd^\alp\sgm_{\alp\dalp}^\mu+\bar{D}_{\dalp}\Lmd^\mu = 0. 
\ee
The most general solution to these conditions can be parametrized by 
\be
 \Lmd^\alp = -\frac{1}{4}\bar{D}^2L^\alp, \;\;\;\;\;
 \Lmd^\mu = -i\sgm^\mu_{\alp\dalp}\bar{D}^{\dalp}L^\alp-\Omg^\mu, 
\ee
where $L^\alp$ is a general complex spinor superfield 
and $\Omg^\mu$ is a chiral superfield. 
In terms of $L^\alp$ and $\Omg^\mu$, 
the transformation of a chiral superfield~$\Phi$ is rewritten as\footnote{
We keep both $L^\alp$ and $\Omg^\mu$ as the transformation parameters 
while the latter is set to zero in Ref.~\cite{Linch:2002wg}. 
In our formulation, 
the degrees of freedom in $\Omg^\mu$ will be eliminated 
by the constraints~(\ref{constraint:Omg}) or absorbed into $\tl{\xi}_{\rm I}$ 
in (\ref{tl:parameters}). }
\bea
 \dlt\Phi \eql -\frac{1}{4}\bar{D}^2L^\alp D_\alp\Phi
 -\brkt{i\sgm^\mu_{\alp\dalp}\bar{D}^{\dalp}L^\alp+\Omg^\mu}\der_\mu\Phi 
 \nonumber\\
 \eql -\frac{1}{4}\bar{D}^2\brkt{L^\alp D_\alp\Phi}
 -\Omg^\mu\der_\mu\Phi.  \label{dlt:Phi}
\eea

Similarly, we can find the transformation acting on 
an anti-chiral superfield~$\bar{\Phi}$ as 
\bea
 \dlt\bar{\Phi} \eql -\frac{1}{4}D^2\bar{L}_{\dalp}\bar{D}^{\dalp}\bar{\Phi}
 -\brkt{i\sgm_{\alp\dalp}^\mu D^\alp\bar{L}^{\dalp}+\bar{\Omg}^\mu}
 \der_\mu\bar{\Phi} \nonumber\\
 \eql -\frac{1}{4}D^2\brkt{\bar{L}_{\dalp}\bar{D}^{\dalp}\bar{\Phi}}
 -\bar{\Omg}^\mu\der_\mu\bar{\Phi}. 
 \label{dlt:bPhi}
\eea
This preserves the anti-chiral condition~$D_\alp\bar{\Phi}=0$. 

Next we consider the transformation of a product of a chiral and 
an anti-chiral superfields~$\Phi_1$ and $\bar{\Phi}_2$. 
In order to define the transformation acting on the product 
that is consistent with (\ref{dlt:Phi}) and (\ref{dlt:bPhi}), 
%In order for the transformation~$\dlt$ to satisfy the Leibniz rule, 
we introduce a real superfield~$U^\mu$ that transforms 
inhomogeneously as 
\be
 \dlt U^\mu = \frac{1}{2}\sgm_{\alp\dalp}^\mu\brkt{
 \bar{D}^{\dalp}L^\alp-D^\alp\bar{L}^{\dalp}}
 -\frac{i}{2}\brkt{\Omg^\mu-\bar{\Omg}^\mu},  \label{dlt:U}
\ee
and insert it into the product as
\be
 \bar{\Phi}_2\Phi_1 \to \cV(\bar{\Phi}_2\Phi_1) 
 \equiv \bar{\Phi}_2\brkt{1+iU^\mu\lrder_\mu}\Phi_1, 
 \label{U_insert1}
\ee
where $A\lrder_\mu B\equiv A\der_\mu B-(\der_\mu A)B$. 
As we will see in the next section, $U^\mu$ contains the SUGRA fields. 
The inserted product~$\cV(\bar{\Phi}_2\Phi_1)$ transforms as 
\bea
 &&\dlt\cV(\bar{\Phi}_2\Phi_1) 
 = \dlt\bar{\Phi}_2\Phi_1+i\bar{\Phi}_2\dlt U^\mu\lrder_\mu\Phi_1
 +\bar{\Phi}_2\dlt\Phi_1 \nonumber\\
 &&= \left\{-\frac{1}{4}\bar{D}^2L^\alp D_\alp
 -\frac{1}{4}D^2\bar{L}_{\dalp}\bar{D}^{\dalp}
 -\frac{i}{2}\sgm_{\alp\dalp}^\mu\brkt{\bar{D}^{\dalp}L^\alp
 +D^\alp\bar{L}^{\dalp}}\der_\mu 
 -\frac{1}{2}\brkt{\Omg^\mu+\bar{\Omg}^\mu}\der_\mu \right\}
 \brkt{\bar{\Phi}_2\Phi_1} \nonumber\\
 &&= \brc{-\frac{1}{4}\bar{D}^2L^\alp D_\alp
 -\frac{1}{2}\brkt{i\sgm_{\alp\dalp}^\mu
 \bar{D}^{\dalp}L^\alp+\Omg^\mu}\der_\mu
 +\hc}\cV(\bar{\Phi}_2\Phi_1). 
 \label{trf:product}
\eea
Here and henceforth, we neglect the $U^\mu$-dependent terms 
in the right-hand sides of the transformation laws 
because they are irrelevant to an invariance of the action 
at the linearized order in $U^\mu$. 
A general superfield has the same transformation law as (\ref{trf:product}). 
Note that this law preserves the reality condition.

\subsection{Superconformal transformations}
It is mentioned in Ref.~\cite{Siegel:1978mj} that 
the transformation~$\dlt$ discussed in the previous subsection 
contain some of the superconformal transformations 
%(\ie, the $\bdm{D}$-, $U(1)_A$-, and $\bdm{S}$-transformations), 
in addition to the super-Poincar\'{e} ones. 
(The 4D $N=1$ superconformal transformations are listed in Table~\ref{4Dsc_trf}.)
%%%%%%%%%%%%%%%%%%%%%%%% Table %%%%%%%%%%%%%%%%%%%%%%%%%%%%%%%%%
\begin{table}[t]
\begin{center}
\begin{tabular}{c|c|c|c} \hline
\rule[-2mm]{0mm}{7mm} $\bdm{P}$ & $\bdm{M}$ & $\bdm{Q}$ & 
\\ \hline
\rule[-2mm]{0mm}{7mm} translation & local Lorentz & SUSY & 
\\ \hline\hline 
\rule[-2mm]{0mm}{7mm} $\bdm{D}$ & $U(1)_A$ & $\bdm{S}$ & $\bdm{K}$ 
\\ \hline 
\rule[-2mm]{0mm}{7mm} dilatation & R-symmetry & conformal SUSY 
& conformal boost \\ \hline
\end{tabular}
\end{center}
\caption{4D $N=1$ superconformal transformations}
\label{4Dsc_trf}
\end{table}
%%%%%%%%%%%%%%%%%%%%%%%%%%%%%%%%%%%%%%%%%%%%%%%%%%%%%%%%%%%%%%%%
However it is unclear how those transformations are related 
to those of Ref.~\cite{Kugo:1982cu}. 
For example, the $\dlt$-transformation law of the conformal compensator superfield 
is essentially different from that of 
a matter chiral superfield~\cite{Linch:2002wg}. 
On the other hand, they transform in the same way 
in the superconformal formulation~\cite{Kugo:1982cu}, 
except for the Weyl weight~$w$ and the chiral weight~$n$, 
which are the charges of $\bdm{D}$ and $U(1)_A$. 

In order to incorporate these weights explicitly 
into the superfield transformations, 
we modify the transformation acting on a general superfield~$\Psi$ as 
\be
 \dsc\Psi \equiv \dlt\Psi+\brkt{w+n}\Lmd\Psi+\brkt{w-n}\bar{\Lmd}\Psi, 
 \label{dsc:Psi}
\ee
where $w$ and $n$ are the Weyl and the chiral weights of $\Psi$,\footnote{
The Weyl and the chiral weights of a superfield denote 
those of the lowest component in the superfield. }  
and $\Lmd$ is a chiral superfield to be determined later. 
(See eq.(\ref{expr:Lmd}).)
This modified transformation~$\dsc$ preserves the (anti-) 
chiral condition or the reality condition, 
and satisfy the Leibniz rule 
since both weights are additive quantum numbers. 
Because $w=n$ and $w=-n$ for a chiral and an anti-chiral 
superfields~$\Phi$ and $\bar{\Phi}$, 
(\ref{dlt:Phi}) and (\ref{dlt:bPhi}) are modified as 
\bea
 \dsc\Phi \eql \brc{-\frac{1}{4}\bar{D}^2L^\alp D_\alp
 -\brkt{i\sgm_{\alp\dalp}^\mu\bar{D}^{\dalp}L^\alp+\Omg^\mu}\der_\mu
 +2w\Lmd}\Phi, \nonumber\\
 \dsc\bar{\Phi} \eql \brc{-\frac{1}{4}D^2\bar{L}_{\dalp}\bar{D}^{\dalp}
 -\brkt{i\sgm_{\alp\dalp}^\mu D^\alp\bar{L}^{\dalp}+\bar{\Omg}^\mu}\der_\mu
 +2w\bar{\Lmd}}\bar{\Phi}.  \label{trf:chiral}
\eea
A real general superfield~$V$ (\ie, $n=0$) transforms as 
\be
 \dsc V = \brc{-\frac{1}{4}\bar{D}^2L^\alp D_\alp
 -\frac{1}{2}\brkt{i\sgm_{\alp\dalp}^\mu\bar{D}^{\dalp}L^\alp+\Omg^\mu}\der_\mu
 +w\Lmd+\hc}V. \label{trf:general}
\ee
The transformation of $U^\mu$ does not change from (\ref{dlt:U}), 
\be
 \dsc U^\mu = \frac{1}{2}\sgm^\mu_{\alp\dalp}
 \brkt{\bar{D}^{\dalp}L^\alp-D^\alp\bar{L}^{\dalp}}
 -\frac{i}{2}\brkt{\Omg^\mu-\bar{\Omg}^\mu}.  \label{dlt:U2}
\ee

The manner of inserting the connection superfield~$U^\mu$ in (\ref{U_insert1}) 
is generalized to a product of 
arbitrary superfields~$\Psi_1,\Psi_2,\cdots,\Psi_n$ as 
\be
 \Psi_1\Psi_2\cdots\Psi_n \to \cV(\Psi_1\Psi_2\cdots\Psi_n)
 \equiv \brkt{1+iU^\mu\hat{\der}_\mu}\brkt{\Psi_1\Psi_2\cdots\Psi_n}, 
 \label{def:cV}
\ee
where 
\be
 \hat{\der}_\mu \equiv \begin{cases} \der_\mu & (\mbox{on chiral superfields}) \\
 0 & (\mbox{on general superfields}) \\
 -\der_\mu & (\mbox{on anti-chiral superfields}) \end{cases}
\ee
Then the transformation~$\dsc$ acts on this product as 
\bea
 \dsc\cV(\Psi_1\Psi_2\cdots\Psi_n)
 \eql \left\{-\frac{1}{4}\bar{D}^2L^\alp D_\alp
 -\frac{1}{2}\brkt{i\sgm_{\alp\dalp}^\mu\bar{D}^{\dalp}L^\alp+\Omg^\mu}\der_\mu
 +\hc \right. \nonumber\\
 &&\left.\hspace{5mm}
 +\brkt{w+n}\Lmd+\brkt{w-n}\bar{\Lmd} \right\}\cV\brkt{\Psi_1\Psi_2\cdots\Psi_n}, 
 \label{dsc:cV}
\eea
where $w$ and $n$ are the Weyl and the chiral weights 
for the product~$\Psi_1\Psi_2\cdots\Psi_n$. 
Notice that $\cV$ defined in (\ref{def:cV}) is regarded as 
an embedding into a general superfield. 
%All the embedded superfields are transformed by 
%a common linear operator in (\ref{dsc:cV}). 

\section{Identification of component fields} \label{comp:identify}
\subsection{SUGRA multiplet and transformation parameters}
Now we will see the transformation laws for component fields. 
First we consider the connection superfields~$U_\mu$, 
whose components are defined as 
\be
 U_\mu = u_\mu+\tht\chi^U_\mu+\bar{\tht}\bar{\chi}^U_\mu
 +\tht^2 a_\mu+\bar{\tht}^2\bar{a}_\mu
 +\brkt{\tht\sgm^\nu\bar{\tht}}\tl{e}_{\nu\mu}
 +\bar{\tht}^2\brkt{\tht\tpsi_\mu}
 +\tht^2\brkt{\bar{\tht}\bar{\tpsi}_\mu}
 +\tht^2\bar{\tht}^2d_\mu, 
\ee
where $u_\mu$, $\tl{e}_{\nu\mu}$ and $d_\mu$ are real. 
Note that $\tl{e}_{\nu\mu}$ is neither symmetric nor anti-symmetric 
for the indices. 
The transformation parameter superfields are expanded as 
\bea
 L_\alp \eql l_\alp+\tht_\alp v+\brkt{\sgm^{\mu\nu}\tht}_\alp w_{\mu\nu}
 -\frac{1}{2}\brkt{\sgm^\mu\bar{\tht}}_\alp\xi_\mu
 +\tht^2\zeta_\alp+\bar{\tht}^2\ep_\alp
 +\tht_\alp\brkt{\eta_\mu\sgm^\mu\bar{\tht}} \nonumber\\
 &&+\frac{1}{2}\bar{\tht}^2\tht_\alp\vph
 -\frac{1}{2}\bar{\tht}^2\brkt{\sgm^{\mu\nu}\tht}_\alp\lmd_{\mu\nu}
 +\tht^2\brkt{\sgm^\mu\bar{\tht}}_\alp\kp_\mu
 -2\tht^2\bar{\tht}^2\rho_\alp, \nonumber\\
 \Omg^\mu \eql \omg^\mu+\tht\zeta_\Omg^\mu+\tht^2 F_\Omg^\mu
 -i\brkt{\tht\sgm^\nu\bar{\tht}}\der_\nu\omg^\mu
 -\frac{i}{2}\tht^2\brkt{\bar{\tht}\bar{\sgm}^\nu\der_\nu\zeta^\mu_\Omg}
 -\frac{1}{4}\tht^2\bar{\tht}^2\Box_4\omg^\mu, 
\eea
where $\Box_4\equiv \eta^{\mu\nu}\der_\mu\der_\nu$, and 
$w_{\mu\nu}$ and $\lmd_{\mu\nu}$ are real anti-symmetric, 
while the others are complex. 
The transformation laws for the component fields 
of $U^\mu$ are read off from (\ref{dlt:U2}) as 
\bea
 \dsc u^\mu \eql \xi_{\rm R}^\mu+\omg_{\rm I}^\mu, \nonumber\\
 \dsc\chi^{U\mu}_\alp \eql -\brkt{\sgm^\mu\bar{\ep}
 -\frac{1}{2}\sgm^\mu\bar{\sgm}^\nu\eta_\nu
 -\frac{i}{2}\sgm^\nu\bar{\sgm}^\mu\der_\nu l
 +\frac{i}{2}\zeta^\mu_\Omg}_\alp, \nonumber\\
 \dsc a^\mu \eql -\kp^\mu+\frac{i}{2}\der^\mu v
 -\frac{i}{2}\der_\nu w^{\nu\mu}
 -\frac{1}{4}\ep^{\mu\nu\rho\tau}\der_\nu w_{\rho\tau}
 -\frac{i}{2}F_\Omg^\mu, \nonumber\\
 \dsc\tl{e}_\nu^{\;\;\mu} \eql -\dlt_\nu^{\;\;\mu}\vph_{\rm R}
 +\lmd^\mu_{\;\;\nu}+\frac{1}{2}\brkt{\der^\mu\xi_{{\rm I}\nu}
 +\der_\nu\xi_{\rm I}^\mu-\dlt_\nu^{\;\;\mu}\der_\rho\xi_{\rm I}^\rho
 +\ep^\mu_{\;\;\nu\rho\tau}\der^\rho\xi_{\rm R}^\tau}
 -\der_\nu\omg_{\rm R}^\mu, \nonumber\\
 \dsc\tpsi_\alp^\mu \eql \brkt{2\sgm^\mu\bar{\rho}
 +\frac{i}{2}\sgm^\nu\bar{\sgm}^\mu\der_\nu\ep
 -\frac{i}{2}\sgm^\nu\der^\mu\bar{\eta}_\nu
 +\frac{1}{4}\sgm^\nu\der_\nu\bar{\zeta}^\mu_\Omg}_\alp, \nonumber\\
 \dsc d^\mu \eql -\frac{1}{2}\der^\mu\vph_{\rm I}
 +\frac{1}{4}\ep^{\mu\nu\rho\tau}\der_\nu\lmd_{\rho\tau}
 -\frac{1}{4}\Box_4\omg_{\rm I}^\mu,  
\eea
where the subscript~R and I denote the real and imaginary parts, respectively. 
By using the freedom of $\Omg^\mu$, we can set 
\be
 u^\mu = \chi^{U\mu}_\alp = a^\mu = 0.  \label{U:WZlike_gauge}
\ee
This is analogous to the Wess-Zumino gauge for a gauge superfield. 
This gauge is preserved if the transformation parameters satisfy 
the following relations. 
\bea
 \omg_{\rm I}^\mu \eql -\xi^\mu_{\rm R}, \nonumber\\
 \zeta_{\Omg\alp}^\mu \eql \brkt{2i\sgm^\mu\bar{\ep}
 -i\sgm^\mu\bar{\sgm}^\nu\eta_\nu
 +\sgm^\nu\bar{\sgm}^\mu\der_\nu l}_\alp, \nonumber\\
 F_\Omg^\mu \eql 2i\kp^\mu+\der^\mu v-\der_\nu w^{\nu\mu}
 +\frac{i}{2}\ep^{\mu\nu\rho\tau}\der_\nu w_{\rho\tau}. 
 \label{constraint:Omg}
\eea
We further impose an additional condition, 
\be
 \xi_{\rm R}^\mu = 0. 
\ee
Then the transformation laws for the residual symmetries reduce to 
\bea
 \dsc\tl{e}_\nu^{\;\;\mu} \eql -\dlt_\nu^{\;\;\mu}\tl{\vph}_{\rm R}
 +\tl{\lmd}^\mu_{\;\;\nu}+\der_\nu\tl{\xi}_{\rm I}^\mu, \nonumber\\
 \dsc\tpsi_\alp^\mu \eql \brkt{2\sgm^\mu\bar{\tl{\rho}}
 +i\sgm^\nu\bar{\sgm}^\mu\der_\nu\ep}_\alp, \nonumber\\
 \dsc d^\mu \eql -\frac{1}{2}\der^\mu\vph_{\rm I}
 +\frac{1}{4}\ep^{\mu\nu\rho\tau}\der_\nu\tl{\lmd}_{\rho\tau}, 
 \label{trf:Weyl}
\eea
where 
\bea
 \tl{\vph}_{\rm R} \defa \vph_{\rm R}
 +\frac{1}{2}\der_\mu\xi^\mu_{\rm I}, \nonumber\\
 \tl{\lmd}_{\mu\nu} \defa \lmd_{\mu\nu}+\frac{1}{2}
 \brkt{\der_\mu\xi_{{\rm I}\nu}-\der_\nu\xi_{{\rm I}\mu}}, \nonumber\\
 \tl{\xi}_{\rm I}^\mu \defa \xi^\mu_{\rm I}-\omg_{\rm R}^\mu, 
 \nonumber\\
 \tl{\rho}^\alp \defa \rho^\alp+\frac{i}{8}
 \brkt{\der_\nu\eta_\mu\sgm^\mu\bar{\sgm}^\nu}^\alp
 +\frac{1}{8}\Box_4 l^\alp.  \label{tl:parameters}
\eea
%Note that component transformations for all the superfields are 
%expressed only by $\tl{\vph}_{\rm R}$, $\vph_{\rm I}$, $\tl{\lmd}_{\mu\nu}$, 
%$\tl{\xi}_{\rm I}^\mu$, $\ep^\alp$ and $\tl{\rho}^\alp$ because 
%the superfield coefficients in (\ref{trf:chiral}) and (\ref{trf:general}) 
%depend only on these parameters, 
%as shown in (\ref{comp:DL1}) and (\ref{comp:DL2}). 
In fact, we can identify $\tl{\xi}_{{\rm I}\mu}$, $\ep_\alp$, 
$\tl{\lmd}_{\mu\nu}$, $\tl{\vph}_{\rm R}$, $-\frac{2}{3}\vph_{\rm I}$ 
and $\tl{\rho}_\alp$ with the transformation parameters for 
$\bdm{P}$, $\bdm{Q}$, $\bdm{M}$, $\bdm{D}$, $U(1)_A$ and 
$\bdm{S}$, respectively. 
They are also expressed as 
\bea
 \tl{\xi}_{\rm I}^\mu \eql 
 \left.-\Re\brkt{i\sgm_{\alp\dalp}^\mu\bar{D}^{\dalp}L^\alp
 +\Omg^\mu}\right|_0, \;\;\;\;\;
 \ep^\alp = \left.-\frac{1}{4}\bar{D}^2L^\alp\right|_0, \nonumber\\
 \tl{\lmd}_{\mu\nu} \eql 
 \left.-\frac{1}{2}\Re\brc{\brkt{\sgm_{\mu\nu}}^{\;\;\alp}_\bt
 D_\alp\bar{D}^2L^\bt}\right|_0, \;\;\;\;\;
 \tl{\vph}_{\rm R} = \left.\Re\brkt{\frac{1}{4}D^\alp\bar{D}^2L_\alp}\right|_0, 
 \nonumber\\
 -\frac{2}{3}\vph_{\rm I} \eql \left.\Im\brkt{-\frac{1}{6}D^\alp\bar{D}^2L_\alp}
 \right|_0, \;\;\;\;\;
 \tl{\rho}^\alp = \left.-\frac{1}{32}D^2\bar{D}^2L^\alp\right|_0, 
\eea
where the symbol~$|_0$ denotes the lowest component of a superfield. 

With the above identification of the parameters, 
the transformations in (\ref{trf:Weyl}) agree with 
those for the Weyl multiplet in Ref.~\cite{Kugo:2002js}, 
which corresponds to the SUGRA multiplet,  
if we specify the components of $U^\mu$ as 
\bea
 \tl{e}_\nu^{\;\;\mu} \eql e_\nu^{\;\;\underline{\mu}}-\dlt_\nu^{\;\;\mu}, 
 \nonumber\\
 \tpsi^\mu_\alp \eql i\brkt{\sgm^\nu\bar{\sgm}^\mu\psi_\nu}_\alp, 
 \nonumber\\
 d^\mu \eql \frac{3}{4}A^\mu
 -\frac{1}{4}\ep^{\mu\nu\rho\tau}\der_\nu \tl{e}_{\rho\tau}, 
 \label{comp:Weyl}
\eea
where $e_\nu^{\;\;\underline{\mu}}$, $\psi_\mu$ and $A_\mu$ are 
the vierbein, the gravitino and the $U(1)_A$ gauge field.\footnote{
The $U(1)_A$ gauge field~$A_\mu$ is an auxiliary field~\cite{Kugo:1982cu}. } 
Namely, $\tl{e}_\nu^{\;\;\mu}$ is the fluctuation of the vierbein since  
$\vev{e_\nu^{\;\;\underline{\mu}}}=\dlt_\nu^{\;\;\mu}$, and 
the (linearized) transformation laws of $\psi_\mu$ and $A_\mu$ are 
given by~\cite{Kugo:2002js}  
\be
 \dsc\psi_{\mu\alp} = \der_\mu\ep_\alp
 +i\brkt{\sgm_\mu\bar{\tl{\rho}}}_\alp, \;\;\;\;\;
 \dsc A_\mu = -\frac{2}{3}\der_\mu\vph_{\rm I}. 
\ee

In the subsequent subsections, we compare the transformation laws 
for component fields in each superfield with those in the superconformal 
formulation~\cite{Kugo:1982cu}, and identify the component fields. 
The transformation laws in Ref.~\cite{Kugo:1982cu} are compactly 
summarized in Sec.~3 of Ref.~\cite{Kugo:2002js}. 
Hence we basically use the notations of Ref.~\cite{Kugo:2002js} 
as the component fields in each multiplet.

\subsection{Chiral multiplet} \label{ID:chiral}
Now we consider the transformation laws of a chiral superfield~$\Phi$. 
In this subsection, we work in the chiral coordinate~$y^\mu\equiv 
x^\mu-i\tht\sgm^\mu\bar{\tht}$. 
(Recall our definition of $\bar{D}_{\dalp}$ in (\ref{def:DbD}).) 
Then, it is expanded as 
\be
 \Phi = \phi+\tht\chi+\tht^2 F. 
% +i\brkt{\tht\sgm^\mu\bar{\tht}}\der_\mu\phi
% -\frac{i}{2}\tht^2\brkt{\bar{\tht}\bar{\sgm}^\mu\der_\mu\chi}
% -\frac{1}{4}\tht^2\bar{\tht}^2\Box_4\phi. 
  \label{comp:chiral}
\ee
Focusing on terms proportional to the Weyl weight 
in the transformation laws in Ref.~\cite{Kugo:2002js}, 
the chiral superfield~$\Lmd$ in (\ref{trf:chiral}) is identified. 
We find that $\Lmd$ cannot be expressed by only $L^\alp$ and $\Omg^\mu$. 
A choice of $\Lmd=-\frac{1}{24}\bar{D}^2D^\alp L_\alp$ reproduces 
the correct $U(1)_A$ transformation, 
but there are extra terms 
for other superconformal transformations.  
%compared to the correct ones in Ref.~\cite{Kugo:1982cu}. 
Fortunately such extra terms are summarized in the form of 
a chiral superfield. 
Thus there exists a choice of $\Lmd$ that realizes the correct 
superconformal transformations for a chiral multiplet. 
It is given by 
\be
 \Lmd = -\frac{1}{24}\brkt{\bar{D}^2D^\alp L_\alp+4\Xi}, 
 \label{expr:Lmd}
\ee
where 
\be
 \Xi \equiv \brkt{-4\tl{\vph}_{\rm R}+\der_\mu\xi_{\rm I}^\mu}
 +8\tht\brkt{\tl{\rho}-\frac{i}{8}\sgm^\nu\bar{\sgm}^\mu\der_\nu\eta_\mu
 +\frac{1}{8}\Box_4 l}+2i\tht^2\der_\mu\brkt{\kp^\mu-\frac{i}{2}\der^\mu v}. 
\ee
Then, the transformation laws of the component fields are read off as 
\bea
 \dsc\phi \eql \tl{\xi}_{\rm I}^\mu\der_\mu\phi+\ep\chi
 +w\tl{\vph}_{\rm R}\phi-\frac{iw}{3}\vph_{\rm I}\phi, \nonumber\\
 \dsc\chi_\alp \eql \tl{\xi}_{\rm I}^\mu\der_\mu\chi_\alp
 +\frac{1}{2}\tl{\lmd}_{\mu\nu}\brkt{\sgm^{\mu\nu}\chi}_\alp
 +2\ep_\alp F-2i\brkt{\sgm^\mu\bar{\ep}}_\alp\der_\mu\phi \nonumber\\
 &&+\brkt{w+\frac{1}{2}}\tl{\vph}_{\rm R}\chi_\alp
 -\frac{i}{3}\brkt{w-\frac{3}{2}}\vph_{\rm I}\chi_\alp
 -4w\tl{\rho}_\alp\phi, \nonumber\\
 \dsc F \eql \tl{\xi}_{\rm I}^\mu\der_\mu F
 -i\bar{\ep}\bar{\sgm}^\mu\der_\mu\chi+\brkt{w+1}\tl{\vph}_{\rm R}F
 -\frac{i}{3}\brkt{w-3}\vph_{\rm I}F+2\brkt{w-1}\tl{\rho}\chi.  
 \label{trf:comp:chiral}
\eea
These transformations agree with those in Ref.~\cite{Kugo:2002js}.

\ignore{
To be exact, the signs of the fermionic bilinear terms are opposite 
to the ones in Ref.~\cite{Kugo:2002js}. 
This discrepancy stems from 
the definition of the spinor covariant derivatives~$D_\alp$ 
and $\bar{D}_{\dalp}$ in Ref.~\cite{Wess:1992cp}. 
The (global) SUSY generators~$Q_\alp$ and $\bar{Q}_{\dalp}$ defined there, 
which anticommute with $D_\alp$ and $\bar{D}_{\dalp}$, satisfy the anticommutation 
relation~$\brc{Q_\alp,\bar{Q}_{\dalp}}=2i\sgm^\mu_{\alp\dalp}\der_\mu$, 
which leads to the wrong sign if we identify $P_\mu=-i\der_\mu$. 
(See Chapter IV of Ref.~\cite{Wess:1992cp}.) 
In fact, we can obtain the transformation laws with the correct signs 
by changing the definition of $D_\alp$ and $\bar{D}_{\dalp}$ as 
\be
 D_\alp \equiv\frac{\der}{\der\tht^\alp}-i\brkt{\sgm^\mu\bar{\tht}}_\alp\der_\mu, 
 \;\;\;\;\;
 \bar{D}_{\dalp} \equiv -\frac{\der}{\der\bar{\tht}^{\dalp}}
 +i\brkt{\tht\sgm^\mu}_{\dalp}\der_\mu,  \label{redefine:DbD}
\ee
which satisfy $\brc{D_\alp,\bar{D}_{\dalp}}=2i\sgm^\mu_{\alp\dalp}\der_\mu$, 
and flipping signs of some fields. 
We summarize the necessary flip of the signs in Sec.~\ref{summary}. 
In practice, the resultant changes will appear only as
the sign flips of all the fermionic bilinear terms. 
Thus we will flip those signs when we compare our results 
with the expressions in Ref.~\cite{Kugo:2002js} in the following. }

Let us comment on a chiral superfield~$\Phi$ 
in the full superspace integral~$\int d^4\tht$. 
Unlike the global SUSY case, 
moving the bases from the chiral coordinate~$y^\mu$ 
to the original one~$x^\mu$ is not enough. 
In fact, $\Phi$ must appear in the form of 
\be
 \cV(\Phi)=\brkt{1+iU^\mu\der_\mu}\Phi. \label{def:cVPhi}
\ee
This is regarded as the embedding of the chiral multiplet 
into a general multiplet as mentioned below (\ref{dsc:cV}). 
The embedded superfield has the following components. 
\bea
 \cV(\Phi) \eql \phi+\tht\chi+\tht^2F
 -i\brkt{\tht\sgm^\mu\bar{\tht}}\brkt{e^{-1}}_\mu^{\;\;\nu}\der_\nu\phi
 -\frac{i}{2}\tht^2\brc{\bar{\tht}\bar{\sgm}^\mu\brkt{e^{-1}}_\mu^{\;\;\nu}
 \der_\nu\chi-2\brkt{\bar{\tht}\bar{\tpsi}^\mu}\der_\mu\phi} \nonumber\\
 &&+i\bar{\tht}^2\brkt{\tht\tpsi^\mu}\der_\mu\phi
 -\frac{1}{4}\tht^2\bar{\tht}^2\brc{g^{\mu\nu}\der_\mu\der_\nu\phi
 +2i\tpsi^\mu\der_\mu\chi-4id^\mu\der_\mu\phi},  \label{comp:cVPhi}
\eea
where $\brkt{e^{-1}}_\mu^{\;\;\nu}\equiv \dlt_\mu^{\;\;\nu}-\tl{e}_\mu^{\;\;\nu}$ 
and $g^{\mu\nu}\equiv \eta^{\mu\nu}-\tl{e}^{\mu\nu}-\tl{e}^{\nu\mu}$ 
are the inverse matrices of the vierbein and the metric, respectively.

\subsection{Real general multiplet}
Next we consider a real general superfield~$V$. 
From (\ref{trf:general}) with (\ref{expr:Lmd}), 
its transformation law is given by 
\be
 \dsc V = \brc{-\frac{1}{4}\bar{D}^2L^\alp D_\alp
 -\frac{1}{2}\brkt{i\sgm^\mu_{\alp\dalp}\bar{D}^{\dalp}L^\alp+\Omg^\mu}\der_\mu
 -\frac{w}{24}\brkt{\bar{D}^2D^\alp L_\alp+4\Xi}+\hc}V. 
 \label{dsc:V}
\ee
Each component of $V$ is defined as 
\be
 V = C'+i\tht\zeta'-i\bar{\tht}\bar{\zeta}'
 -\tht^2\cH'-\bar{\tht}^2\bar{\cH}'
 -\brkt{\tht\sgm^\mu\bar{\tht}}B'_\mu
 +i\tht^2\brkt{\bar{\tht}\bar{\lmd}'}
 -i\bar{\tht}^2\brkt{\tht\lmd'}
 +\frac{1}{2}\tht^2\bar{\tht}^2D',  \label{comp:real_general}
\ee
where $C'$, $B'_\mu$ and $D'$ are real. 
By expanding (\ref{dsc:V}) in terms of the components, 
we obtain their transformation laws. 
They do not agree with the transformation laws in Ref.~\cite{Kugo:2002js} 
as is. 
To reproduce the latter laws, we need to redefine the components as 
\bea
 C \defa C', \;\;\;\;\;
 \zeta_\alp \equiv \zeta'_\alp, \;\;\;\;\;
 \cH \equiv \cH',  \nonumber\\
 B_\mu \defa B'_\mu+\zeta'\psi_\mu+\bar{\zeta}'\bar{\psi}_\mu
 +\frac{w}{2}C' A_\mu, \nonumber\\
 \lmd_\alp \defa \lmd'_\alp
 +\frac{i}{2}\brc{\sgm^\mu\brkt{e^{-1}}_\mu^{\;\;\nu}
 \der_\nu\bar{\zeta}'}_\alp
 +\brkt{\sgm^\mu\bar{\sgm}^\nu\psi_\mu}_\alp B'_\nu
 +\frac{w}{4}\brkt{\sgm^\mu\bar{\zeta}'}_\alp A_\mu, \nonumber\\
 D \defa D'+\frac{1}{2}g^{\mu\nu}\der_\mu\der_\nu C'
 +\brkt{\bar{\lmd}'\bar{\sgm}^\mu\psi_\mu
 -\frac{i}{2}\der_\nu\zeta'\sgm^\mu\bar{\sgm}^\nu\psi_\mu
 -i\der_\mu\zeta'\psi^\mu
 -\frac{2iw}{3}\zeta'\sgm^{\mu\nu}\der_\nu\psi_\mu+\hc} \nonumber\\
 &&-\brkt{2d^\mu-\frac{w}{2}A^\mu}B'_\mu.  
 \label{Vcomp:redefine}
\eea
The explicit form of $V$ in terms of these redefined components is shown 
in (\ref{expr:comp_vector}). 
Then we obtain 
\bea
 \dsc C \eql \tl{\xi}_{\rm I}^\mu\der_\mu C
 +i\ep\zeta -i\bar{\ep}\bar{\zeta}+w\tl{\vph}_{\rm R}C, \nonumber\\
 \dsc\zeta_\alp \eql \tl{\xi}_{\rm I}^\mu\der_\mu\zeta_\alp
 +\frac{1}{2}\tl{\lmd}_{\mu\nu}\brkt{\sgm^{\mu\nu}\zeta}_\alp
 +2i\ep_\alp\cH+\brkt{\sgm^\mu\bar{\ep}}_\alp 
 \brkt{iB_\mu-\der_\mu C} \nonumber\\
 &&+\brkt{w+\frac{1}{2}}\tl{\vph}_{\rm R}\zeta_\alp
 +\frac{i}{2}\vph_{\rm I}\zeta_\alp+2iw\tl{\rho}_\alp C, \nonumber\\
 \dsc\cH \eql \tl{\xi}_{\rm I}^\mu\der_\mu\cH
 -i\bar{\ep}\bar{\lmd}-\bar{\ep}\bar{\sgm}^\mu\der_\mu\zeta
 +\brkt{w+1}\tl{\vph}_{\rm R}\cH
 +i\vph_{\rm I}\cH-i\brkt{w-2}\tl{\rho}\zeta, \nonumber\\
 \dsc B_\mu \eql \tl{\xi}_{\rm I}^\nu\der_\nu B_\mu
 +\tl{\lmd}_{\mu\nu}B^\nu
 -i\ep\sgm_\mu\bar{\lmd}-i\bar{\ep}\bar{\sgm}_\mu\lmd
 -\ep\der_\mu\zeta-\bar{\ep}\der_\mu\bar{\zeta} \nonumber\\
 &&+\brkt{w+1}\tl{\vph}_{\rm R}B_\mu
 -i\brkt{w+1}\brkt{\tl{\rho}\sgm_\mu\bar{\zeta}
 +\bar{\tl{\rho}}\bar{\sgm}_\mu\zeta}, \nonumber\\
 \dsc\lmd_\alp \eql \tl{\xi}_{\rm I}^\mu\der_\mu\lmd_\alp
 +\frac{1}{2}\tl{\lmd}_{\mu\nu}\brkt{\sgm^{\mu\nu}\lmd}_\alp
 +i\ep_\alp D-\brkt{\sgm^{\mu\nu}\ep}_\alp
 \brkt{\der_\mu B_\nu-\der_\nu B_\mu}
 +\brkt{w+\frac{3}{2}}\tl{\vph}_{\rm R}\lmd_\alp \nonumber\\
 &&+\frac{iw}{2}\der_\mu\tl{\vph}_{\rm R}\brkt{\sgm^\mu\bar{\zeta}}_\alp 
 -\frac{i}{2}\vph_{\rm I}\lmd_\alp
 -iw\brkt{\sgm^\mu\bar{\tl{\rho}}}_\alp B_\mu
 +2iw\tl{\rho}_\alp\bar{\cH}
 +w\brkt{\sgm^\mu\bar{\tl{\rho}}}_\alp\der_\mu C, \nonumber\\
 \dsc D \eql \tl{\xi}_{\rm I}^\mu\der_\mu D
 +2\der_\mu\tl{\lmd}^{\mu\nu}\der_\nu C
 +\brkt{w+2}\tl{\vph}_{\rm R}D
 +w\der_\mu\tl{\vph}_{\rm R}\der^\mu C \nonumber\\
 &&-\brc{\bar{\ep}\bar{\sgm}^\mu\der_\mu\lmd
 +2iw\tl{\rho}\lmd+w\tl{\rho}\sgm^\mu\der_\mu\bar{\zeta}+\hc}. 
 \label{dsc:vct}
\eea
These transformation laws agree with those in Ref.~\cite{Kugo:2002js} 
at the linearized level,\footnote{
The complex scalar~$\cH$ should be understood as 
$\frac{1}{2}\brkt{H+iK}$ in the notation of Ref.~\cite{Kugo:2002js}. } 
except for the following two points. 
\begin{itemize}
\item The second term in $\dsc D$ is absent in Ref.~\cite{Kugo:2002js}. 
 However, this term is harmless when we consider the invariance of 
 the action because $D$ is the highest component 
 and this term is a total derivative. 
 
\item The terms proportional to $\der_\mu\tl{\vph}_{\rm R}$ 
 in $\dsc\lmd_\alp$ and $\dsc D$ also seem to be extra terms 
 that are absent in Ref.~\cite{Kugo:2002js}, at first glance. 
 Actually, they indicate that the $\bdm{D}$-gauge field~$b_\mu$ is set to zero 
 in our linearized SUGRA. 
 Its superconformal transformation (at the linearized level) is 
 \be
  \dsc b_\mu = \der_\mu\tl{\vph}_{\rm R}-2\xi_{K\mu}, 
 \ee
 where $\xi_{K\mu}$ is the transformation parameter for $\bdm{K}$. 
 Thus, keeping the condition~$b_\mu=0$ requires 
 $\xi_{K\mu}=\frac{1}{2}\der_\mu\tl{\vph}_{\rm R}$. 
 In fact, after the replacement:~$\der_\mu\tl{\vph}_{\rm R}\to 2\xi_{K\mu}$, 
 (\ref{dsc:vct}) reproduces the correct transformations. 
\end{itemize}

\subsection{Gauge multiplet} \label{gauge_multiplet}
Here we consider a gauge multiplet, which corresponds to 
a real general multiplet with $w=0$.\footnote{
We consider an abelian gauge multiplet for simplicity. 
An extension to the nonabelian case is straightforward. } 
From (\ref{Vcomp:redefine}), 
such a real general superfield~$V$ is expressed as 
\bea
 V \eql C+i\tht\zeta-i\bar{\tht}\bar{\zeta}
 -\tht^2\cH-\bar{\tht}^2\bar{\cH}
 -\brkt{\tht\sgm^\mu\bar{\tht}}\brkt{e^{-1}}_\mu^{\;\;\nu}
 \hat{B}_\nu \nonumber\\
 &&+i\tht^2\bar{\tht}\brc{\bar{\lmd}
 -\frac{i}{2}\bar{\sgm}^\mu\brkt{e^{-1}}_\mu^{\;\;\nu}
 \der_\nu\zeta-i\bar{\tpsi}_\mu\hat{B}^\mu} 
 -i\bar{\tht}^2\tht\brc{\lmd-\frac{i}{2}\sgm^\mu
 \brkt{e^{-1}}_\mu^{\;\;\nu}\der_\nu\bar{\zeta}
 +i\tpsi_\mu\hat{B}^\mu} \nonumber\\
 &&+\frac{1}{2}\tht^2\bar{\tht}^2\left\{
 D-\frac{1}{2}g^{\mu\nu}\der_\mu\der_\nu C
 +\brkt{-\frac{i}{2}\bar{\lmd}\bar{\sgm}^\mu\tpsi_\mu
 +\der_\mu\zeta\tpsi^\mu+\hc}+2d^\mu\hat{B}_\mu \right\}, 
\eea
where 
\be
 \hat{B}_\mu \equiv \brkt{\dlt_\mu^{\;\;\nu}+\tl{e}_\mu^{\;\;\nu}}B_\nu
 -\zeta\psi_\mu-\bar{\zeta}\bar{\psi}_\mu, 
\ee
is interpreted as a gauge field. 
This definition of the gauge field is consistent with 
that of Ref.~\cite{Kugo:1982cu}. 

The gauge transformation can be defined just in a similar way 
to the global SUSY case as 
\be
 V \to V+\cV(\Sgm)+\cV(\bar{\Sgm}),  \label{gauge_trf}
\ee
where $\Sgm=\phi_\Sgm+\tht\chi_\Sgm+\tht^2 F_\Sgm$ 
(in the coordinate~$y^\mu=x^\mu-i\tht\sgm^\mu\bar{\tht}$) is a chiral superfield. 
Note that $\Sgm$ must be embedded into a general multiplet by $\cV$ 
in order to be added to $V$. 
We can move to the Wess-Zumino gauge by choosing $\Sgm$ as 
\be
 \Re\phi_\Sgm = -\frac{1}{2}C, \;\;\;\;\;
 \chi_{\Sgm\alp} = -i\zeta_\alp, \;\;\;\;\;
 F_\Sgm = \cH. 
\ee
In this gauge, $V$ is written as 
\bea
 V_{\rm WZ} \eql -\brkt{\tht\sgm^\mu\bar{\tht}}\brkt{e^{-1}}_\mu^{\;\;\nu}
 \hat{B}'_\nu+i\tht^2\bar{\tht}\brkt{\bar{\lmd}-i\bar{\tpsi}^\mu\hat{B}'_\mu}
 -i\bar{\tht}^2\tht\brkt{\lmd+i\tpsi^\mu\hat{B}'_\mu} \nonumber\\
 &&+\frac{1}{2}\tht^2\bar{\tht}^2\brc{D
 +\brkt{-\frac{i}{2}\bar{\lmd}\bar{\sgm}^\mu\tpsi_\mu+\hc}
 +2d^\mu\hat{B}'_\mu},  \label{WZ_gauge}
\eea
where $\hat{B}'_\mu\equiv\hat{B}_\mu-2\der_\mu\Im\phi_\Sgm$ 
is the gauge-transformed gauge field. 
We can move to this gauge only when $w=0$. 
The set of the components $[\hat{B}'_\mu,\lmd_\alp,D]$ form 
a gauge multiplet in Ref.~\cite{Kugo:1982cu,Kugo:2002js}. 

Next we construct a field strength superfield~$\cW_\alp$ 
that is gauge-invariant from the gauge superfield~$V$. 
A naive definition of $\cW_\alp$, 
\be
 \cW^{\rm naive}_\alp \equiv -\frac{1}{4}\bar{D}^2D_\alp V, 
\ee
is not invariant under (\ref{gauge_trf}). 
If we define 
\be
 X \equiv \brkt{1+\frac{1}{4}U^\mu\bar{\sgm}_\mu^{\dbt\bt}
 \sbk{D_\bt,\bar{D}_{\dbt}}}V, 
\ee
its gauge transformation becomes simpler, 
\be
 X \to X+\Sgm+\bar{\Sgm}. 
\ee
Hence, $-\frac{1}{4}\bar{D}^2D_\alp X$ becomes gauge-invariant. 
However, this is not the only way to construct a gauge-invariant quantity. 
We can define the following quantity by adding the second term 
that is also gauge-invariant. 
\be
 \hat{\cW}_\alp \equiv -\frac{1}{4}\bar{D}^2D_\alp X
 +c\bar{D}^2\brkt{U^\mu\sgm_\mu^{\dbt\bt}D_\alp D_\bt\bar{D}_{\dbt}V}, 
 \label{def:hatcW}
\ee
where $c$ is a constant to be determined later. 
The second term in (\ref{def:hatcW}) does not contribute to the lowest component, 
and we find that 
\be
 \hat{\cW}_\alp = -i\brkt{\lmd
 +\frac{1}{2}\tl{e}_\mu^{\;\;\nu}\sgm^\mu\bar{\sgm}_\nu\lmd}_\alp
 +\cO(\tht).  
\ee
This indicates that we have to multiply $\hat{\cW}_\alp$ 
by a superfield~$Z_\alp^{\;\;\bt}=
\dlt_\alp^{\;\;\bt}-\frac{1}{2}\tl{e}_\mu^{\;\;\nu}
\brkt{\sgm^\mu\bar{\sgm}_\nu}_\alp^{\;\;\bt}+\cO(\tht)$ in order to obtain 
the desired field strength superfield whose lowest component is 
$-i\lmd_\alp$. 
The higher components of $Z_\alp^{\;\;\bt}$ and the constant~$c$ 
in (\ref{def:hatcW}) are determined so that 
$\cW_\alp\equiv Z_\alp^{\;\;\bt}\hat{\cW}_\bt$ has the correct components. 
%Then we can determine the constant~$c$ as $c=\frac{1}{8}$ 
%so that the correct bosonic components of $\cW_\alp$ are obtained. 
%Finally, the fermionic component of $Z_\alp^{\;\;\bt}$ is determined 
%by the condition that the correct fermionic components of $\cW_\alp$ 
%are obtained. 
The result is 
\bea
 \cW_\alp \eql -\frac{1}{4}Z_\alp^{\;\;\bt}\bar{D}^2
 \brc{D_\bt\brkt{V+\frac{1}{4}U^\mu\bar{\sgm}_\mu^{\dgm\gm}
 \sbk{D_\gm,\bar{D}_{\dgm}}V}
 -\frac{1}{2}U^\mu\bar{\sgm}_\mu^{\dgm\gm}D_\bt D_\gm\bar{D}_{\dgm}V} \nonumber\\
 \eql -\frac{1}{4}Z_\alp^{\;\;\bt}\bar{D}^2
 \brc{D_\bt V+\frac{1}{4}D_\bt U^\mu\bar{\sgm}_\mu^{\dgm\gm}
 \sbk{D_\gm,\bar{D}_{\dgm}}V-iU^\mu\der_\mu D_\bt V}, \nonumber\\
 Z_\alp^{\;\;\bt} \defa \dlt_\alp^{\;\;\bt}
 -\frac{1}{2}\tl{e}_\mu^{\;\;\nu}\brkt{\sgm^\mu\bar{\sgm}_\nu}_\alp^{\;\;\bt}
 -\brkt{\sgm^\mu\bar{\tpsi}_\mu}_\alp \tht^\bt,  \label{def:cW}
\eea
where $Z_\alp^{\;\;\bt}$ is expressed in the chiral coordinate~$y^\mu$. 
Each component of $\cW_\alp$ is calculated as 
\bea
 \cW_\alp \eql -i\lmd_\alp+\tht_\alp D
 +i\brkt{\sgm^{\mu\nu}\tht}_\alp\brkt{e^{-1}}_\mu^{\;\;\rho}
 \brkt{e^{-1}}_\nu^{\;\;\tau}\hat{F}_{\rho\tau}
 -\tht^2\brc{\sgm^\mu\brkt{e^{-1}}_\mu^{\;\;\nu}\cD_\nu\bar{\lmd}}_\alp, 
\eea
where 
\bea
 \hat{F}_{\mu\nu} \defa \der_\mu\hat{B}_\nu-\der_\nu\hat{B}_\mu
 +i\brkt{\psi_\mu\sgm_\nu\bar{\lmd}-\psi_\nu\sgm_\mu\bar{\lmd}}
 +i\brkt{\bar{\psi}_\mu\bar{\sgm}_\nu\lmd
 -\bar{\psi}_\nu\bar{\sgm}_\mu\lmd}, \nonumber\\
 \brkt{\cD_\mu\bar{\lmd}}^{\dalp} \defa 
 \brc{\brkt{\der_\mu-\frac{1}{2}\omg_\mu^{\;\;\nu\rho}\sgm_{\nu\rho}
 +\frac{3i}{4}A_\mu}\bar{\lmd}}^{\dalp}
 +\brkt{\bar{\sgm}^{\nu\rho}\bar{\psi}_\mu}^{\dalp}\hat{F}_{\nu\rho}
 +i\bar{\psi}_\mu^{\dalp}D. 
 \label{def:F_Dlmd}
\eea
Here $\omg_\mu^{\;\;\nu\rho}$ is the spin connection, and expressed 
at the linearized level as 
\bea
 \omg_\mu^{\;\;\nu\rho} \eql 
 -\frac{1}{2}\der_\mu\brkt{\tl{e}^{\nu\rho}-\tl{e}^{\rho\nu}}
 +\frac{1}{2}\brc{\der^\nu\brkt{\tl{e}_\mu^{\;\;\rho}
 +\tl{e}^\rho_{\;\;\mu}}
 -\der^\rho\brkt{\tl{e}_\mu^{\;\;\nu}+\tl{e}^\nu_{\;\;\mu}}}. 
\eea
The field strength~$\hat{F}_{\mu\nu}$ and 
the covariant derivative~$\cD_\mu\bar{\lmd}$ defined in (\ref{def:F_Dlmd}) 
agree with those in Ref.~\cite{Kugo:2002js}. 
We have used the identification~(\ref{comp:Weyl}).

\section{Action formulae}  \label{Action_fml}
\subsection{$\bdm{F}$-term action formula}
First we consider the $F$-term invariant action, 
which consists of only chiral multiplets. 
It reduces to the chiral superspace integral in the global SUSY limit, 
\be
 S_F^{\rm gl}[W] \equiv \int\dr^4x \int\dr^2\tht\;W+\hc, 
\ee
where $W$ is a chiral superfield, which is referred to as the superpotential. 
From (\ref{trf:chiral}) with (\ref{expr:Lmd}), this transforms as 
\bea
 \dsc S_F^{\rm gl}[W] \eql \int\dr^4x \int\dr^2\tht\brc{
 -\frac{1}{4}\bar{D}^2\brkt{L^\alp D_\alp W}-\Omg^\mu\der_\mu W
 -\frac{1}{4}\brkt{\bar{D}^2D^\alp L_\alp+4\Xi}W}+\hc \nonumber\\
 \eql \int\dr^4x\int\dr^2\tht\;\brc{
 -\frac{1}{4}\bar{D}^2D^\alp\brkt{L_\alp W}-\Omg^\mu\der_\mu W
 -\Xi W}+\hc \nonumber\\
 \eql \int\dr^4x\sbk{\int\dr^4\tht\;D^\alp\brkt{L_\alp W}
 -\int\dr^2\tht\;\brkt{\Xi-\der_\mu\Omg^\mu}W}+\hc. 
\eea
We have assumed that $w=n=3$ for $W$. 
In the last equation, we performed the partial integrals. 
In order to make the action invariant, 
we introduce a chiral superfield~$\tl{\cE}$ 
whose transformation law is given by 
\bea
 \dsc\tl{\cE} \eql \Xi-\der_\mu\Omg^\mu \nonumber\\
 \eql \brkt{-4\tl{\vph}_{\rm R}+\der_\mu\tl{\xi}_{\rm I}^\mu}
 +8\tht\brkt{\tl{\rho}-\frac{i}{4}\sgm^\mu\der_\mu\bar{\ep}},  \label{trf:Y}
\eea
in the coordinate~$y^\mu$, and modify the action formula as 
\be
 S_F[W] \equiv \int\dr^4x\int\dr^2\tht\;\brkt{1+\tl{\cE}}W+\hc.  \label{def:S_F}
\ee
From (\ref{trf:Weyl}) and (\ref{trf:Y}), we identify $\tl{\cE}$ as 
\be
 \tl{\cE} = \tl{e}_\mu^{\;\;\mu}-\tht\sgm^\mu\bar{\tpsi}_\mu 
 = \tl{e}_\mu^{\;\;\mu}-2i\tht\sgm^\mu\bar{\psi}_\mu. \label{def:Y}
\ee
Therefore, the factor~$(1+\tl{\cE})$ in (\ref{def:S_F}) 
corresponds to the chiral density multiplet in Ref.~\cite{Wess:1992cp}.\footnote{ 
Note that $\det\brkt{e_\mu^{\;\;\underline{\nu}}}=1+\tl{e}_\mu^{\;\;\mu}$ 
at the linearized level.}

%Due to the modification~(\ref{def:S_F}), 
The action~$S_F[W]$ is now invariant 
under $\dsc$ at the linearized level. 
Note that it is invariant only when the Weyl weight of $W$ is 3. 
This is consistent with the $F$-term action formula 
in Ref.~\cite{Kugo:1982cu}, which is shown in (\ref{F-formula}) 
in our notations. 
We can explicitly see that (\ref{def:S_F}) reproduces (\ref{F-formula}).

\subsection{$\bdm{D}$-term action formula} \label{D_action_fml}
Next we consider the $D$-term invariant action. 
It reduces to the full superspace integral 
in the global SUSY limit,\footnote{
The factor 2 is necessary to match the normalization of 
the $D$-term action formula in Ref.~\cite{Kugo:1982cu}. }
\be
 S_D^{\rm gl}[K] \equiv 2\int\dr^4x\int\dr^4\tht\;K, 
\ee
where $K$ is a real general superfield, which is referred to 
as the K\"{a}hler potential. 
From (\ref{trf:general}) with (\ref{expr:Lmd}), this transforms as 
\bea
 \dsc S_D^{\rm gl}[K] \eql 2\int\dr^4x\int\dr^4\tht\left\{
 -\frac{1}{4}\bar{D}^2L^\alp D_\alp
 -\frac{1}{2}\brkt{i\sgm_{\alp\dalp}^\mu\bar{D}^{\dalp}L^\alp+\Omg^\mu}\der_\mu 
 \right. \nonumber\\
 &&\left.\hspace{25mm}
 -\frac{w}{24}\brkt{\bar{D}^2D^\alp L_\alp+4\Xi}+\hc \right\}K \\
 \eql 2\int\dr^4x\int\dr^4\tht\brc{\frac{6-w}{24}\bar{D}^2D^\alp L_\alp
 -\frac{1}{2}\brkt{i\sgm_{\alp\dalp}^\mu\der_\mu\bar{D}^{\dalp}L^\alp
 -\der_\mu\Omg^\mu}-\frac{w}{6}\Xi+\hc}K, \nonumber
\eea
where $w$ is the Weyl weight of $K$. 
We have performed the partial integral in the second equality. 
Here we define a real scalar superfield~$\tl{E}_1$ 
from the connection superfield~$U^\mu$ as  
\be
 \tl{E}_1 \equiv \frac{1}{4}\bar{\sgm}^{\dalp\alp}_\mu
 \sbk{D_\alp,\bar{D}_{\dalp}}U^\mu.  \label{def:tlE}
\ee
Then it transforms as 
\be
 \dsc\tl{E}_1 = -\frac{1}{2}\bar{D}^2D^\alp L_\alp
 +\frac{3i}{2}\sgm^\mu_{\alp\dalp}\der_\mu\bar{D}^{\dalp}L^\alp
 -\frac{1}{2}\der_\mu\Omg^\mu+\hc.  \label{dsc:E1}
\ee
By using $\tl{E}_1$ and $\tl{\cE}$ defined in (\ref{def:Y}), 
we modify the action formula as 
\be
 S_D[K] \equiv 2\int\dr^4x\int\dr^4\tht\;
 \brc{1+\frac{1}{3}\brkt{\tl{E}_1+\tl{\cE}+\bar{\tl{\cE}}}}K, 
 \label{def:S_D}
\ee
so that its transformation becomes 
\be
 \dsc S_D[K] = 2\int\dr^4x\int\dr^4\tht\;\brc{
 \frac{2-w}{24}\bar{D}^2D^\alp L_\alp+\frac{2-w}{6}\Xi+\hc}K.  
\ee
Therefore, $S_D[K]$ is now $\dsc$-invariant when $w=2$, 
and can be identified with the $D$-term action formula 
in Ref.~\cite{Kugo:1982cu}, which is shown in (\ref{D-formula}) 
in our notations.  
Since the prefactor of $K$ in (\ref{def:S_D}) is expanded 
as (see (\ref{expr:density})) 
\be
 1+\frac{1}{3}\brkt{\tl{E}_1+\tl{\cE}+\bar{\tl{\cE}}} 
 = 1+\tl{e}_\mu^{\;\;\mu}+\cO(\tht^2), 
\ee
this corresponds to the density multiplet in the full superspace~\cite{Wess:1992cp}. 
We can explicitly show that (\ref{def:S_D}) reproduces (\ref{D-formula}), 
except for the kinetic terms for the SUGRA fields, which will be discussed 
in Sec.~\ref{SUGRA_kin}.

\subsection{Absorption of chiral density superfield}
Notice that the ``chiral density superfield''~$\tl{\cE}$ defined in (\ref{def:Y}) 
is redundant because it cannot be expressed in terms of $U^\mu$ and 
the SUGRA fields are already contained in the latter. 
In fact, we can eliminate $\tl{\cE}$ 
from the action formulae~(\ref{def:S_F}) and (\ref{def:S_D}) 
by the following superfield redefinition. 
\bea
 \hat{\Phi} \defa \brkt{1+\frac{w}{3}\tl{\cE}}\Phi, \nonumber\\
 \hat{V} \defa \brc{1+\frac{w}{6}\brkt{\tl{\cE}+\bar{\tl{\cE}}}}V, 
 \label{redefinition}
\eea
where $\Phi$ and $V$ are a chiral and a real general superfields, 
respectively. 
The redefined superfields transform as 
\bea
 \dsc\hat{\Phi} \eql \brc{-\frac{1}{4}\bar{D}^2L^\alp D_\alp
 -\brkt{i\sgm_{\alp\dalp}^\mu\bar{D}^{\dalp}L^\alp+\Omg^\mu}\der_\mu
 -\frac{w}{12}\brkt{\bar{D}^2D^\alp L_\alp+4\der_\mu\Omg^\mu}}
 \hat{\Phi}, \label{modified:dsc} \\
 \dsc\hat{V} \eql \brc{-\frac{1}{4}\bar{D}^2L^\alp D_\alp
 -\frac{1}{2}\brkt{i\sgm_{\alp\dalp}^\mu\bar{D}^{\dalp}L^\alp+\Omg^\mu}\der_\mu
 -\frac{w}{24}\brkt{\bar{D}^2D^\alp L_\alp+4\der_\mu\Omg^\mu}+\hc}
 \hat{V}.  \nonumber
\eea
Now the transformations are expressed only in terms of $L^\alp$ and $\Omg^\mu$. 
In terms of these redefined superfields, the action formulae are expressed as 
\bea
 S_F[W] \eql 
 \int\dr^4x\int\dr^2\tht\;\hat{W}+\hc, \nonumber\\
 S_D\sbk{K} 
 \eql 2\int\dr^4x\int\dr^4\tht\;\brkt{1+\frac{1}{3}\tl{E}_1}\hat{K}  \nonumber\\
 \eql 2\int\dr^4x\int\dr^4\tht\;\brkt{1+\frac{1}{12}
 \bar{\sgm}_\mu^{\dalp\alp}\sbk{D_\alp,\bar{D}_{\dalp}}U^\mu}\hat{K}.  
 \label{redefined_actions}
\eea

As for the gauge kinetic term, the $\tl{\cE}$-dependence automatically cancels 
with another redundant superfield~$Z_\alp^{\;\;\bt}$ in (\ref{def:cW}) 
because 
\bea
 \cW_\alp \eql Z_\alp^{\;\;\bt}\hat{\cW}_\bt
 = \hat{\cW}_\alp-\frac{1}{2}\tl{e}_\mu^{\;\;\nu}
 \brkt{\sgm^\mu\bar{\sgm}_\nu\hat{\cW}}_\alp
 -\brkt{\tht\hat{\cW}}\brkt{\sgm^\mu\bar{\tpsi}_\mu}_\alp, \nonumber\\
 \cW^\alp\cW_\alp \eql \brc{1-\tl{e}_\mu^{\;\;\mu}+\tht\sgm^\mu\bar{\tpsi}_\mu}
 \hat{\cW}^\alp\hat{\cW}_\alp 
 = \brkt{1-\tl{\cE}}\hat{\cW}^\alp\hat{\cW}_\alp. 
\eea
Thus we obtain 
\bea
 S^{\rm gauge}_{\rm kin}[V] \defa S_F\sbk{-\frac{1}{4}\cW^\alp\cW_\alp} 
 \nonumber\\
 \eql \int\dr^4x\int\dr^2\tht\;\brkt{1+\tl{\cE}}
 \brkt{-\frac{1}{4}\cW^\alp\cW_\alp}+\hc  \nonumber\\
 \eql \int\dr^4x\int\dr^2\tht\;\brkt{-\frac{1}{4}
 \hat{\cW}^\alp\hat{\cW}_\alp}+\hc, 
\eea
where 
\be
 \hat{\cW}_\alp = 
 -\frac{1}{4}\bar{D}^2\brkt{
 D_\alp\hat{V}+\frac{1}{4}D_\alp U^\mu\bar{\sgm}_\mu^{\dbt\bt}
 \sbk{D_\bt,\bar{D}_{\dbt}}\hat{V}-iU^\mu\der_\mu D_\alp\hat{V}}. 
 \label{def:hatcW2}
\ee
Note that $\hat{V}=V$ since the Weyl weight of the gauge superfield is zero.

\subsection{Kinetic terms for SUGRA fields} \label{SUGRA_kin}
%So far, we have neglected terms beyond the linear order in the SUGRA fields. 
Here we discuss the kinetic terms for the SUGRA superfield~$U^\mu$. 
In the superconformal formulation, the corresponding terms 
are contained in the $D$-term action formula 
in Ref.~\cite{Kugo:1982cu} as follows. 
(See eq.(\ref{D-formula}).)
\be
 S_D\sbk{\Omg} = e\sbk{D+\cdots+\frac{C}{3}\brc{R(\omg)
 +4\ep^{\mu\nu\rho\tau}\brkt{\psi_\mu\sgm_\tau\der_\nu\bar{\psi}_\rho+\hc}}}, 
 \label{expr:S_D:comp}
\ee
where $e$ is the determinant of the vierbein, 
$R(\omg)$ is the scalar curvature constructed from the spin connection, 
and $\Omg=[C,\cdots,D]$ is a real general multiplet 
with the Weyl weight 2. 
The Einstein-Hilbert term is obtained by imposing the $\bdm{D}$-gauge 
fixing condition, $C=-\frac{3}{2}$.\footnote{ 
We have taken the unit of the Planck mass, $M_{\rm Pl}=1$. } 
Since the above kinetic terms are quadratic in the SUGRA fields, 
it seems that 
we need to extend the $D$-term action formula in (\ref{redefined_actions}) as 
\be
 S_D[\Omg] = 2\int\dr^4x\int\dr^4\tht\;\brc{1+\frac{1}{3}\brkt{\tl{E}_1
 +\tl{E}_2}}\hat{\Omg}, \label{S_D2}
\ee
where $\tl{E}_2$ is quadratic in $U^\mu$. 
The quadratic part~$\tl{E}_2$ is specified by requiring 
the invariance of the action up to linear order in the SUGRA fields. 
However, information on higher order corrections to (\ref{dsc:E1}) 
and (\ref{modified:dsc}) in the SUGRA fields is necessary for this procedure, 
which is beyond the linearized SUGRA. 
Fortunately, it is possible to extend (\ref{redefined_actions}) 
to include the kinetic terms for $U^\mu$ 
without information on the higher order corrections, 
as we will explain below. 

Recall that the SUGRA kinetic terms are proportional to 
the lowest component of $\hat{\Omg}$, which will be set to 
a constant~$\Omg_0=-\frac{3}{2}$ after the $\bdm{D}$-gauge fixing. 
Thus we expand $\hat{\Omg}$ as 
\be
 \hat{\Omg} = \Omg_0+\tl{\hat{\Omg}}, 
\ee
and focus on $\tl{E}_2\Omg_0$ out of $\tl{E}_2\hat{\Omg}$ in (\ref{S_D2}). 
Namely, we consider 
\bea
 S_D[\Omg] \eql 2\int\dr^4x\int\dr^4\tht\brc{
 \brkt{1+\frac{1}{3}\tl{E}_1}\brkt{\Omg_0+\tl{\hat{\Omg}}}
 +\frac{1}{3}\tl{E}_2\Omg_0} \nonumber\\
 \eql 2\int\dr^4x\int\dr^4\tht\brc{\frac{1}{3}\tl{E}_2\Omg_0
 +\brkt{1+\frac{1}{3}\tl{E}_1}\tl{\hat{\Omg}}},  \label{S_D3}
\eea
as an extension of (\ref{redefined_actions}). 
We have used a fact that $\tl{E}_1$ is a total derivative 
at the second equality. 
For example, when $\hat{\Omg}$ is given by 
\be
 \hat{\Omg} = -\frac{3}{2}|\hat{\Phi}_C|^2e^{-\hat{K}/3},  
\ee
where $\hat{K}$ is quadratic in matter fields and 
$\hat{\Phi}_C=1+\cdots$ is a chiral compensator superfield,  
the action~(\ref{S_D3}) is written as 
\be
 S_D[\Omg] = \int\dr^4x\int\dr^4\tht\brc{-\tl{E}_2
 +\brkt{1+\frac{1}{3}\tl{E}_1}\brkt{\hat{K}+\cdots}}, 
\ee
where the ellipsis denotes higher order terms. 

We require the action~(\ref{S_D3}) to be invariant up to linear order 
in the SUGRA fields for $\Omg_0$-dependent terms 
while up to zeroth order in the SUGRA fields 
for $\tl{\hat{\Omg}}$-dependent terms. 
Up to this order, 
\be
 \int\dr^4x\int\dr^4\tht\brc{\frac{1}{3}\brkt{\dsc\tl{E}_1}\tl{\hat{\Omg}}
 +\dsc\tl{\hat{\Omg}}} = 0, 
\ee
as shown in Sec.~\ref{D_action_fml}. 
Hence the variation of (\ref{S_D3}) becomes 
\be
 \dsc S_D[\Omg] = 2\int\dr^4x\int\dr^4\tht\brc{
 \frac{1}{3}\brkt{\dsc\tl{E}_2}\Omg_0+\frac{1}{3}\tl{E}_1\dsc\tl{\hat{\Omg}}}. 
 \label{dsc:S_D}
\ee
In order to discuss the invariance of the action up to the order 
under consideration, 
we only need a field-independent part of $\dsc\tl{\hat{\Omg}}$. 
From (\ref{modified:dsc}), it is read off as 
\be
 \dsc\tl{\hat{\Omg}} = \dsc\hat{\Omg} = -\frac{\Omg_0}{12}
 \brkt{\bar{D}^2D^\alp L_\alp+4\der_\mu\Omg^\mu+\hc}+\cdots, 
\ee
where the ellipsis denotes field-dependent terms. 
We have used that the Weyl weight of $\Omg$ is 2. 
Since the above field-independent part of $\dsc\tl{\hat{\Omg}}$ 
is not affected by including higher order corrections 
in the SUGRA fields, the variation~(\ref{dsc:S_D}) becomes 
\be
 \dsc S_D[\Omg] = 
 2\int\dr^4x\int\dr^4\tht\;\frac{\Omg_0}{3}\brc{
 \dsc\tl{E}_2
 -\frac{1}{12}\tl{E}_1\brkt{\bar{D}^2D^\alp L_\alp+4\der_\mu\Omg^\mu+\hc}}.
\ee
After some calculations, we can show that 
\be
 \dsc\brc{-\frac{1}{8}U_\mu D^\alp\bar{D}^2D_\alp U^\mu
 +\frac{1}{3}\tl{E}_1^2-\brkt{\der_\mu U^\mu}^2} = 
 \frac{1}{6}\tl{E}_1\brkt{\bar{D}^2D^\alp L_\alp+4\der_\mu\Omg^\mu+\hc}, 
\ee
where total derivatives are dropped. 
Therefore, $\tl{E}_2$ is identified as 
\be
 \tl{E}_2 = -\frac{1}{16}U_\mu D^\alp\bar{D}^2D_\alp U^\mu
 +\frac{1}{6}\tl{E}_1^2-\frac{1}{2}\brkt{\der_\mu U^\mu}^2. 
 \label{def:R_2}
\ee
This has a similar form to the counterpart of Ref.~\cite{Linch:2002wg}. 
The first term of the second line in (\ref{S_D3}) with (\ref{def:R_2}) 
is the kinetic terms for $U^\mu$.

\ignore{
Note that $\tl{\hat{\Omg}}$ transforms inhomogeneously because $\dsc\Omg_0=0$. 
From (\ref{modified:dsc}), its transformation law is read off as 
\bea
 \dsc\tl{\hat{\Omg}} \eql -\frac{\Omg_0}{12}
 \brkt{\bar{D}^2D^\alp L_\alp+4\der_\mu\Omg^\mu} \\
 &&+\brc{-\frac{1}{4}\bar{D}^2L^\alp D_\alp
 -\frac{1}{2}\brkt{i\sgm^\mu_{\alp\dalp}\bar{D}^{\dalp}L^\alp+\Omg^\mu}\der_\mu
 -\frac{1}{12}\brkt{\bar{D}^2D^\alp L_\alp+4\der_\mu\Omg^\mu}}\tl{\hat{\Omg}}
 +\hc. \nonumber
\eea
We have used that the Weyl weight of $\Omg$ is 2. 
Since we have taken into account quadratic terms 
in the SUGRA fields in the action, we should extend this transformation 
by including higher order corrections. 
First, the redefinition of a real general superfield in (\ref{redefinition}) is 
extended as 
\be
 \hat{\Omg} \equiv \brkt{1+Y_1+Y_2}\Omg
 = \brkt{1+Y_1+Y_2}\brkt{\Omg_0+\tl{\Omg}}, \label{extend:hatOmg}
\ee
where $Y_1\equiv\frac{1}{3}\brkt{\tl{\cE}+\bar{\tl{\cE}}}$ 
and $Y_2$ is quadratic in the SUGRA fields. 
The transformation laws~(\ref{trf:Y}) and (\ref{trf:general}) 
with (\ref{expr:Lmd}) are also extended as 
\bea
 \dsc\tl{\cE} \eql \Xi-\der_\mu\Omg^\mu+\cX_1, \nonumber\\
 \dsc\tl{\Omg} \eql \brc{-\frac{1}{12}\brkt{\bar{D}^2D^\alp L_\alp+4\Xi}
 +\cY'_1+\hc}\Omg_0+\brkt{\mbox{$\tl{\Omg}$-dependent terms}}+\cdots \nonumber\\
 \eql \brc{-\frac{1}{12}\brkt{\bar{D}^2D^\alp L_\alp+4\Xi}
 +\cY_1+\hc}\Omg_0+\brkt{\mbox{$\tl{\hat{\Omg}}$-dependent terms}}+\cdots, 
\eea
where $\cX_1$, $\cY'_1$ and $\cY_1$ are linear in the SUGRA fields, 
and the ellipsis denotes terms beyond linear order in the SUGRA fields. 
Then, the transformation law of $\tl{\hat{\Omg}}$ is calculated as  
\bea
 \dsc\tl{\hat{\Omg}} \eql \dsc\hat{\Omg} 
 = \brkt{\dsc Y_1+\dsc Y_2}\brkt{\Omg_0+\tl{\Omg}}
 +\brkt{1+Y_1+Y_2}\dsc\tl{\Omg} \nonumber\\
 \eql \brkt{\dsc Y_1+\dsc Y_2}\brkt{1+Y_1+Y_2}^{-1}\Omg_0
 +\brkt{1+Y_1+Y_2}\dsc\tl{\Omg}
 +\brkt{\mbox{$\tl{\hat{\Omg}}$-dependent terms}} \nonumber\\
% \eql \brc{\frac{1}{3}\brkt{\Xi-\der_\mu\Omg^\mu+\cX_1+\hc}+\dsc Y_2}\Omg_0 
% \nonumber\\
% &&
% +\brkt{1+Y_1+Y_2}\brc{-\frac{1}{12}\brkt{\bar{D}^2D^\alp L_\alp+4\Xi}
% +\cY_1+\hc}\Omg_0+\cdots \nonumber\\
 \eql \Omg_0\left\{-\frac{1}{12}\brkt{\bar{D}^2D^\alp L_\alp+4\der_\mu\Omg^\mu+\hc}
 +\dsc Y_2+\brkt{\frac{1}{3}\cX_1+\cY_1+\hc} \right. \nonumber\\
 &&\left.\hspace{10mm}
 -\frac{1}{12}Y_1\brkt{\bar{D}^2D^\alp L_\alp+4\der_\mu\Omg^\mu+\hc}\right\}
 +\brkt{\mbox{$\tl{\hat{\Omg}}$-dependent terms}}+\cdots, 
\eea
where the ellipsis denotes terms quadratic in the SUGRA fields. 
Thus if we choose $Y_2$ such that  
\be
 \dsc Y_2 = -\brkt{\frac{1}{3}\cX_1+\cY_1+\hc}
 +\frac{1}{12}Y_1\brkt{\bar{D}^2D^\alp L_\alp+4\der_\mu\Omg^\mu+\hc}, 
\ee
we obtain 
\be
 \dsc\tl{\hat{\Omg}} = -\frac{\Omg_0}{12}
 \brkt{\bar{D}^2D^\alp L_\alp+4\der_\mu\Omg^\mu}
 +\brkt{\mbox{$\tl{\hat{\Omg}}$-dependent terms}}+\cdots, 
\ee
where the ellipsis denotes terms quadratic in the SUGRA fields. 
Now we consider the variation of the action~(\ref{S_D3}) 
in order to specify $\tl{E}_2$. 
We require the action to be invariant up to linear order in the SUGRA fields 
for $\Omg_0$-dependent terms while up to zeroth order in the SUGRA fields 
for $\tl{\hat{\Omg}}$-dependent terms.  
Up to this order, 
\be
 \int\dr^4x\int\dr^4\tht\brc{\frac{1}{3}\brkt{\dsc\tl{E}_1}\tl{\hat{\Omg}}
 +\dsc\tl{\hat{\Omg}}} = 0, 
\ee
as shown in Sec.~\ref{D_action_fml}. 
Hence the variation of (\ref{S_D3}) becomes 
\bea
 \dsc S_D[\Omg] \eql 2\int\dr^4x\int\dr^4\tht\brc{
 \frac{1}{3}\brkt{\dsc\tl{E}_2}\Omg_0+\frac{1}{3}\tl{E}_1\dsc\tl{\hat{\Omg}}} 
 \nonumber\\
 \eql 2\int\dr^4x\int\dr^4\tht\;\frac{\Omg_0}{3}\brc{
 \dsc\tl{E}_2
 -\frac{1}{12}\tl{E}_1\brkt{\bar{D}^2D^\alp L_\alp+4\der_\mu\Omg^\mu+\hc}}.
\eea
After some calculations, we can show that 
\be
 \dsc\brc{-\frac{1}{8}U_\mu D^\alp\bar{D}^2D_\alp U^\mu
 +\frac{1}{3}\tl{E}_1^2-\brkt{\der_\mu U^\mu}^2} = 
 \frac{1}{6}\tl{E}_1\brkt{\bar{D}^2D^\alp L_\alp+4\der_\mu\Omg^\mu+\hc}, 
\ee
where total derivatives are dropped. 
Therefore, $\tl{E}_2$ is identified as 
\be
 \tl{E}_2 = -\frac{1}{16}U_\mu D^\alp\bar{D}^2D_\alp U^\mu
 +\frac{1}{6}\tl{E}_1^2-\frac{1}{2}\brkt{\der_\mu U^\mu}^2. 
 \label{def:R_2}
\ee
This has a similar form to the counterpart of Ref.~\cite{Linch:2002wg}. 
The first term of the second line in (\ref{S_D3}) with (\ref{def:R_2}) 
is the kinetic terms for $U^\mu$. 
}

Finally we comment on the relation of the superfield action~(\ref{S_D3}) 
to the component expression~(\ref{expr:S_D:comp}). 
In order to reproduce the quadratic part of 
the SUSY Einstein-Hilbert terms~$\cL_{\rm quad}^{\rm SG}$ 
in (\ref{L^SG_quad}), 
we also need to count the SUGRA fields contained 
in the redefined superfields, in addition to the kinetic terms for $U^\mu$. 
By including higher order corrections in the SUGRA fields, 
the redefinition of a real general superfield in (\ref{redefinition}) is 
extended as 
\be
 \hat{\Omg} \equiv \brkt{1+Y_1+Y_2}\Omg
 = \brkt{1+Y_1+Y_2}\brkt{\Omg_0+\tl{\Omg}}, \label{extend:hatOmg}
\ee
where $Y_1\equiv\frac{1}{3}\brkt{\tl{\cE}+\bar{\tl{\cE}}}$ 
and $Y_2$ is quadratic in the SUGRA fields. 
Thus, from (\ref{S_D3}) and (\ref{extend:hatOmg}), the $\Omg_0$-dependent part 
of the action is expressed as  
\be
 S_D[\Omg] = 2\int\dr^4x\int\dr^4\tht\;\brkt{\frac{1}{3}\tl{E}_2
 +\frac{1}{3}\tl{E}_1Y_1+Y_2}\Omg_0+\cdots, 
\ee
where the ellipsis denotes terms beyond quadratic order 
in the SUGRA fields or depending on the matter fields. 
This corresponds to $\cL^{\rm SG}_{\rm quad}$ in (\ref{L^SG_quad}).

\section{Summary} \label{summary}
We have modified the 4D $N=1$ linearized SUGRA, 
and provided a complete identification of component fields 
in each superfield with fields in the superconformal formulation 
of SUGRA developed in Ref.~\cite{Kugo:1982cu}. 
The results of our work makes it possible to use both formulations 
in a complementary manner. 

In our modified linearized SUGRA, 
(anti-) chiral superfields and real general superfields 
should be understood as the redefined ones defined 
in (\ref{redefinition}) whose components are identified with 
the fields in Ref.~\cite{Kugo:1982cu} through 
(\ref{comp:chiral}), (\ref{comp:real_general}), 
(\ref{Vcomp:redefine}) and (\ref{def:Y}). 
The components of the connection superfield~$U^\mu$ are
identified with the SUGRA fields in the Weyl-multiplet 
as (\ref{comp:Weyl}). 
The invariant action formulae are expressed 
in terms of the redefined superfields as 
\bea
 S_F[W] \eql \int\dr^4x\int\dr^2\tht\;\hat{W}+\hc, \nonumber\\
 S_D[\Omg] \eql 2\int\dr^4x\int\dr^4\tht\brc{\frac{\Omg_0}{3}\tl{E}_2
 +\brkt{1+\frac{1}{3}\tl{E}_1}\hat{\Omg}}, \nonumber\\
 S^{\rm gauge}_{\rm kin}[V] 
% \defa S_F\sbk{-\frac{1}{4}\cW^\alp\cW_\alp}  \nonumber\\
 \eql \int\dr^4x\int\dr^2\tht\brkt{
 -\frac{1}{4}\hat{\cW}^\alp\hat{\cW}_\alp}+\hc, 
\eea
where $\tl{E}_1$ and $\tl{E}_2$ are defined 
in (\ref{def:tlE}) and (\ref{def:R_2}), and
the field strength superfield~$\hat{\cW}_\alp$ 
is defined in (\ref{def:hatcW2}). 
$\Omg_0$ is a constant part of $\hat{\Omg}$, 
which is set to $-3/2$ for the canonically normalized SUGRA kinetic terms.  
In the $D$-term action formula, a chiral multiplet~$\hat{\Phi}$ 
must be embedded into a general multiplet. 
Such embedding is provided at the linearized order in the SUGRA fields by 
\be
 \cV(\hat{\Phi}) \equiv \brkt{1+iU^\mu\der_\mu}\hat{\Phi}. 
\ee
The gauge transformation of the gauge multiplet~$\hat{V}$ is given by 
\be
 \hat{V} \to \hat{V}+\cV(\hat{\Sgm})+\cV(\bar{\hat{\Sgm}}), 
\ee
where $\hat{\Sgm}$ is a chiral superfield. 
The field strength superfield~$\hat{\cW}_\alp$ is invariant 
under this transformation. 

\ignore{
As mentioned in Sec.~\ref{ID:chiral}, we have to use the spinor 
derivatives~$D_\alp$ and $\bar{D}_{\dalp}$ as (\ref{redefine:DbD}) 
in order to obtain expressions identical to those 
in Refs.~\cite{Kugo:1982cu,Kugo:2002js}. 
This corresponds to the sign flip~$\der_\mu\to -\der_\mu$ 
in all expressions in the text. 
Accordingly some of the fields also need to 
change their signs to match the expressions in Ref.~\cite{Kugo:1982cu,Kugo:2002js}. 
They are $(\chi_\alp,F)$ in a chiral multiplet, 
$(\zeta_\alp,B_\mu,\cH)$ in a real general multiplet, 
and $(\psi_{\mu\alp},A_\mu)$ in the Weyl multiplet.\footnote{
The signs of $\tpsi_\alp$ and $d_\mu$ do not change.}
The transformation parameters for $\bdm{P}$ and $\bdm{S}$ become 
$-\tl{\xi}_{\rm I}$ and $-\tl{\rho}_\alp$, respectively. 
In practice, the resultant changes only appear as the sign flip 
of fermionic bilinear terms in all the expressions. }

Our work will also be useful to discuss higher-dimensional SUGRA. 
When we consider it in the context 
of the brane-world scenario, it is convenient to express the action 
in terms of $N=1$ superfields, keeping only $N=1$ SUSY that remains 
unbroken at low energies manifest. 
The authors of Ref.~\cite{Linch:2002wg} construct 
minimal version of 5D linearized SUGRA along this direction. 
Although their formulation is powerful to calculate 
SUGRA loop contributions and is self-consistent, 
it is not clear how the component fields are related to fields 
in other off-shell formulations of 5D SUGRA. 
Especially, it is obscure how to extend their result 
to more general 5D SUGRA. 
The superconformal formulation of 5D SUGRA has been developed 
in Refs.~\cite{5D_Kugo,Kugo:2002js}, 
and its $N=1$ superfield description was provided 
in Ref.~\cite{Paccetti:2004ri,Abe:2004ar}. 
Hence, by combining these results with our current work, 
it is possible to obtain the linearized SUGRA formulation of 
{\it general} 5D SUGRA. 
Furthermore, our work is a good starting point to construct 
an $N=1$ description of 6D or higher-dimensional SUGRA 
because the linearized SUGRA formulation is based on the ordinary 
superspace and thus is easier to handle than 
the full supergravity. 
These issues are left for our future works.

\subsection*{Acknowledgements}
This work was supported in part by 
Grant-in-Aid for Young Scientists (B) No.22740187 
from Japan Society for the Promotion of Science.

\appendix

\section{Component expressions} \label{comp_expr}
Here we collect component expressions of some superfields 
appearing in the text. 

The coefficient superfields of the differential operators in (\ref{trf:chiral}) 
and (\ref{trf:general}) are written as 
\bea
 -\frac{1}{4}\bar{D}^2L^\alp \eql 
 \ep^\alp+\frac{1}{2}\tht^\alp\tl{\vph}
 -\frac{1}{2}\brkt{\tht\sgm^{\mu\nu}}^\alp\tl{\lmd}_{\mu\nu}
 -2\tht^2\tl{\rho}^\alp
 -i\brkt{\tht\sgm^\mu\bar{\tht}}\der_\mu\ep^\alp \nonumber\\
 &&-\frac{i}{4}\tht^2\brc{
 \brkt{\bar{\tht}\bar{\sgm}^\mu}^\alp\der_\mu\tl{\vph}
 -\brkt{\bar{\tht}\bar{\sgm}^\rho\sgm^{\mu\nu}}^\alp
 \der_\rho\tl{\lmd}_{\mu\nu}}
 -\frac{1}{4}\tht^2\bar{\tht}^2\Box_4\ep^\alp, \nonumber\\
 i\sgm^\mu_{\alp\dalp}\bar{D}^{\dalp}L^\alp+\Omg^\mu
 \eql -\tl{\xi}_{\rm I}^\mu
 +2i\tht\sgm^\mu\bar{\ep}+2i\bar{\tht}\bar{\sgm}^\mu\ep
 -i\brkt{\tht\sgm_\nu\bar{\tht}}
 \brkt{\eta^{\mu\nu}\tl{\vph}+\tl{\lmd}^{\mu\nu}
 -\der^\nu\tl{\xi}_{\rm I}^\mu+\frac{i}{2}\ep^{\mu\nu\rho\tau}\tl{\lmd}_{\rho\tau}}
 \nonumber\\
 &&-4i\tht^2\brc{\bar{\tht}\bar{\sgm}^\mu\tl{\rho}
 +\frac{i}{4}\brkt{\bar{\tht}\bar{\sgm}^\nu\sgm^\mu\der_\nu\bar{\ep}}}
 -\bar{\tht}^2\brkt{\tht\sgm^\nu\bar{\sgm}^\mu\der_\nu\ep} \nonumber\\
 &&-\frac{1}{2}\tht^2\bar{\tht}^2\der_\nu\brkt{
 \eta^{\mu\nu}\tl{\vph}+\tl{\lmd}^{\mu\nu}
 -\frac{1}{2}\der^\nu\tl{\xi}_{\rm I}^\mu
 +\frac{i}{2}\ep^{\mu\nu\rho\tau}\tl{\lmd}_{\rho\tau}},  \label{comp:DL1}
\eea
where $\tl{\vph}\equiv \tl{\vph}_{\rm R}+i\vph_{\rm I}$, 
$\tl{\lmd}_{\mu\nu}$, $\tl{\xi}_{\rm I}$ 
and $\tl{\rho}^\alp$ are defined in (\ref{tl:parameters}). 
An explicit expression of $\Lmd$ defined in (\ref{expr:Lmd}) is 
\bea
 \Lmd \eql -\frac{1}{24}\brkt{\bar{D}^2D^\alp L_\alp+4\Xi} \nonumber\\
 \eql \frac{1}{2}\brkt{\tl{\vph}_{\rm R}-\frac{i}{3}\vph_{\rm I}}
 -2\tht\tl{\rho}-\frac{i}{2}\brkt{\tht\sgm^\mu\bar{\tht}}
 \der_\mu\brkt{\tl{\vph}_{\rm R}-\frac{i}{3}\vph_{\rm I}} \nonumber\\
 &&+i\tht^2\brkt{\bar{\tht}\bar{\sgm}^\mu\der_\mu\tl{\rho}}
 -\frac{1}{8}\tht^2\bar{\tht}^2\Box_4
 \brkt{\tl{\vph}_{\rm R}-\frac{i}{3}\vph_{\rm I}}.  \label{comp:DL2}
\eea
We have taken the gauge~(\ref{U:WZlike_gauge}). 

For a real general multiplet~$[C,\zeta_\alp,\cH,B_\mu,\lmd_\alp,D]$, 
each component is embedded into a real superfield that transforms 
by a law~(\ref{trf:general}) as 
\bea
 V \eql C+i\tht\zeta-i\bar{\tht}\bar{\zeta}-\tht^2\cH-\bar{\tht}^2\bar{\cH}
 -\brkt{\tht\sgm^\mu\bar{\tht}}
 \brkt{B_\mu-\zeta\psi_\mu-\bar{\zeta}\bar{\psi}_\mu
 -\frac{w}{2}C A_\mu} \nonumber\\
 &&+i\tht^2\bar{\tht}\brc{\bar{\lmd}-\frac{i}{2}\bar{\sgm}^\mu
 \brkt{e^{-1}}_\mu^{\;\;\nu}\der_\nu\zeta
 -\brkt{\bar{\sgm}^\mu\sgm^\nu\bar{\psi}_\mu}B_\nu
 +\frac{w}{4}\brkt{\bar{\sgm}^\mu\zeta}A_\mu} \nonumber\\
 &&-i\bar{\tht}^2\tht\brc{\lmd-\frac{i}{2}\sgm^\mu\brkt{e^{-1}}_\mu^{\;\;\nu}
 \der_\nu\bar{\zeta}-\brkt{\sgm^\mu\bar{\sgm}^\nu\psi_\mu}B_\nu
 -\frac{w}{4}\brkt{\sgm^\mu\bar{\zeta}}A_\mu} \nonumber\\
 &&+\frac{1}{2}\tht^2\bar{\tht}^2\left\{
 D-\frac{1}{2}g^{\mu\nu}\der_\mu\der_\nu C
 -\brkt{\frac{w-3}{2}A^\mu+\frac{1}{2}\ep^{\mu\nu\rho\tau}\der_\nu\tl{e}_{\rho\tau}}
 B_\mu\right. \nonumber\\
 &&\hspace{15mm}\left.
 +\brkt{-\bar{\lmd}\bar{\sgm}^\mu\psi_\mu
 -2i\der_\mu\zeta\sgm^{\mu\nu}\psi_\nu
 +i\der_\mu\zeta\psi^\mu
 +\frac{2iw}{3}\zeta\sgm^{\mu\nu}\der_\nu\psi_\mu+\hc} \right\}. 
 \label{expr:comp_vector}
\eea

The real superfield~$\tl{E}_1$ defined in (\ref{def:tlE}) is expressed as  
\bea
 \tl{E}_1 \eql \frac{1}{4}\bar{\sgm}_\mu^{\dalp\alp}
 \sbk{D_\alp,\bar{D}_{\dalp}}U^\mu \nonumber\\
 \eql \tl{e}_\mu^{\;\;\mu}
 +\tht\sgm^\mu\bar{\tpsi}_\mu-\bar{\tht}\bar{\sgm}^\mu\tpsi_\mu
 +\brkt{\tht\sgm^\mu\bar{\tht}}
 \brkt{2d_\mu-\ep_{\mu\nu\rho\tau}\der^\nu\tl{e}^{\rho\tau}} \nonumber\\
 &&+\frac{i}{2}\bar{\tht}^2\brkt{\tht\sgm^\mu\bar{\sgm}^\nu\der_\nu\tpsi_\mu}
 -\frac{i}{2}\tht^2\brkt{\bar{\tht}\bar{\sgm}^\mu\sgm^\nu\der_\nu\bar{\tpsi}_\mu}
 +\frac{1}{4}\tht^2\bar{\tht}^2
 \brkt{\Box_4\tl{e}_\mu^{\;\;\mu}-2\der_\mu\der_\nu\tl{e}^{\mu\nu}} \nonumber\\
 \eql \tl{e}_\mu^{\;\;\mu}
 +2i\brkt{\tht\sgm^\mu\bar{\psi}_\mu
 +\bar{\tht}\bar{\sgm}^\mu\psi_\mu}
 +\frac{3}{2}\brkt{\tht\sgm^\mu\bar{\tht}}
 \brkt{A_\mu-\ep_{\mu\nu\rho\tau}\der^\nu\tl{e}^{\rho\tau}} \nonumber\\
 &&-2\bar{\tht}^2\brkt{\tht\der_\mu\psi^\mu}
 -2\tht^2\brkt{\bar{\tht}\der_\mu\bar{\psi}^\mu}
 +\frac{1}{4}\tht^2\bar{\tht}^2
 \brkt{\Box_4\tl{e}_\mu^{\;\;\mu}-2\der_\mu\der_\nu\tl{e}^{\mu\nu}}. 
\eea
Thus the density superfield is calculated as 
\bea
 1+\frac{1}{3}\brkt{\tl{E}_1+\tl{\cE}+\bar{\tl{\cE}}} 
 \eql \brkt{1+\tl{e}_\mu^{\;\;\mu}}
 +\frac{1}{2}\brkt{\tht\sgm^\mu\bar{\tht}}
 \brkt{A_\mu-\ep_{\mu\nu\rho\tau}\der^\nu\tl{e}^{\rho\tau}} \nonumber\\
 &&-\frac{1}{3}\bar{\tht}^2\brc{\tht\brkt{2\eta^{\mu\nu}+\sgm^\mu\bar{\sgm}^\nu}
 \der_\mu\psi_\nu}
 -\frac{1}{3}\tht^2\brc{\bar{\tht}\brkt{2\eta^{\mu\nu}+\bar{\sgm}^\mu\sgm^\nu}
 \der_\mu\bar{\psi}_\nu} \nonumber\\
 &&-\frac{1}{12}\tht^2\bar{\tht}^2\brkt{\Box_4\tl{e}_\mu^{\;\;\mu}
 +2\der_\mu\der_\nu\tl{e}^{\mu\nu}}.  \label{expr:density}
\eea

\section{Invariant action formulae} \label{FD_action_fml}
Here we collect the invariant action formulae in Ref.~\cite{Kugo:1982cu} 
in our notations. 
For a chiral multiplet~$\Phi=\sbk{\phi,\chi_\alp,F}$ with 
weight $(w,n)=(3,3)$, the $F$-term action formula is given by
\bea
 S_F[\Phi] \eql = \int\dr^4x\;e\brkt{F-i\bar{\psi}_\mu\bar{\sgm}^\mu\chi
 +\hc+\cdots} \nonumber\\
 \eql \int\dr^4x\brc{\brkt{1+\tl{e}_\mu^{\;\;\mu}}F
 -i\bar{\psi}_\mu\bar{\sgm}^\mu\chi+\hc+\cdots},   \label{F-formula}
\eea
where $e\equiv\det\brkt{e_\mu^{\;\;\underline{\nu}}}$, 
and the ellipsis denotes terms beyond the linear order in the SUGRA fields. 

For a real general multiplet~$\Omg=[C,\zeta_\alp,\cH,B_\mu,\lmd_\alp,D]$ 
with weights~$(w,n)=(2,0)$, the $D$-term action formula is given by
\bea
 S_D[\Omg] \eql \int\dr^4x\;e\left[D-\bar{\psi}_\mu\bar{\sgm}^\mu\lmd
 +\psi_\mu\sgm^\mu\bar{\lmd}
 +\frac{4i}{3}\brkt{\zeta\sgm^{\mu\nu}\der_\mu\psi_\nu
 -\bar{\zeta}\bar{\sgm}^{\mu\nu}\der_\mu\bar{\psi}_\nu} \right. \nonumber\\
 &&\hspace{15mm}\left. 
 +\frac{C}{3}\brc{R(\omg)+4\ep^{\mu\nu\rho\tau}
 \brkt{\psi_\mu\sgm_\tau\der_\nu\bar{\psi}_\rho
 -\bar{\psi}_\mu\bar{\sgm}_\tau\der_\nu\psi_\rho}}
 +\cdots \right] \nonumber\\
 \eql \int\dr^4x\left[\brkt{1+\tl{e}_\mu^{\;\;\mu}}D
 +\brkt{\psi_\mu\sgm^\mu\bar{\lmd}
 +\frac{4i}{3}\zeta\sgm^{\mu\nu}\der_\mu\psi_\nu+\hc} \right. \nonumber\\
 &&\hspace{15mm}\left. 
 +\frac{2\tl{C}}{3}\brkt{\der^\mu\der_\nu\tl{e}_\mu^{\;\;\nu}
 -\Box_4\tl{e}_\mu^{\;\;\mu}}+\cL^{\rm SG}_{\rm quad}+\cdots \right], 
 \label{D-formula}
\eea
where $R(\omg)$ is the scalar curvature constructed 
from the spin connection, $\tl{C}\equiv C-\Omg_0$ 
where $\Omg_0$ is a constant to which $C$ will be set 
by the $\bdm{D}$-gauge fixing. 
Thus $\tl{C}$ will vanish after the $\bdm{D}$-gauge fixing. 
The quadratic part in the SUGRA fields~$\cL^{\rm SG}_{\rm quad}$ is given by 
\bea
 \cL^{\rm SG}_{\rm quad} \eql \frac{\Omg_0}{3}\left\{
 \tl{e}_\mu^{\;\;\mu}\brkt{2\der^\nu\der_\rho\tl{e}_\nu^{\;\;\rho}
 -\Box_4\tl{e}_\nu^{\;\;\nu}}
 -\tl{e}^{\mu\nu}\brkt{\der_\nu\der^\rho\tl{e}_{(\mu\rho)} 
 +\der_\mu\der^\rho\tl{e}_{(\nu\rho)}
 -\Box_4\tl{e}_{(\mu\nu)}} \right. \nonumber\\
 &&\left.\hspace{10mm} 
 +4\ep^{\mu\nu\rho\tau}\brkt{\psi_\mu\sgm_\tau\der_\nu\bar{\psi}_\rho+\hc}
 \right\},   \label{L^SG_quad}
\eea
where 
$\tl{e}_{(\mu\nu)} \equiv \frac{1}{2}\brkt{\tl{e}_{\mu\nu}+\tl{e}_{\nu\mu}}$, 
and the ellipsis denotes terms beyond the linear order in the SUGRA fields.

%%%%%%%%%%%%%%%%%%%%%%%%%%%% References %%%%%%%%%%%%%%%%%%%%%%%%%%%%%%


\begin{thebibliography}{99}
\bibitem{Kaku:1978nz}
  M.~Kaku, P.~K.~Townsend, P.~van Nieuwenhuizen,
  %``Properties of Conformal Supergravity,''
  Phys.\ Rev.\  {\bf D17 } (1978)  3179.
  
\bibitem{Kaku:1978ea}
  M.~Kaku, P.~K.~Townsend,
  %``Poincare Supergravity As Broken Superconformal Gravity,''
  Phys.\ Lett.\  {\bf B76 } (1978)  54.

\bibitem{Ferrara:1978rk}
  S.~Ferrara, M.~T.~Grisaru, P.~van Nieuwenhuizen,
  %``Poincare And Conformal Supergravity Models With Closed Algebras,''
  Nucl.\ Phys.\  {\bf B138 } (1978)  430.

\bibitem{Kugo:1982cu}
  T.~Kugo, S.~Uehara,
  %``Conformal And Poincare Tensor Calculi In N=1 Supergravity,''
  Nucl.\ Phys.\  {\bf B226 } (1983)  49.
  
\bibitem{5D_Kugo}
  T.~Kugo, K.~Ohashi,
  %``Off-shell D = 5 supergravity coupled to matter Yang-Mills system,''
  Prog.\ Theor.\ Phys.\  {\bf 105 } (2001)  323-353. [hep-ph/0010288]; 
  T.~Fujita, T.~Kugo, K.~Ohashi,
  %``Off-shell formulation of supergravity on orbifold,''
  Prog.\ Theor.\ Phys.\  {\bf 106 } (2001)  671-690 
  [hep-th/0106051].

\bibitem{Kugo:2002js}
  T.~Kugo, K.~Ohashi,
  %``Superconformal tensor calculus on orbifold in 5D,''
  Prog.\ Theor.\ Phys.\  {\bf 108 } (2002)  203-228.
  [hep-th/0203276].

\bibitem{Ferrara:1977mv}
  S.~Ferrara, B.~Zumino,
  %``Structure of Conformal Supergravity,''
  Nucl.\ Phys.\  {\bf B134 } (1978)  301. 
  
\bibitem{Siegel:1978mj}
  W.~Siegel, S.~J.~Gates, Jr.,
  %``Superfield Supergravity,''
  Nucl.\ Phys.\  {\bf B147 } (1979)  77.

\bibitem{Linch:2002wg}
  W.~D.~Linch, III, M.~A.~Luty, J.~Phillips,
  %``Five-dimensional supergravity in N=1 superspace,''
  Phys.\ Rev.\  {\bf D68 } (2003)  025008.
  [hep-th/0209060]. 
  
\bibitem{Buchbinder:2003qu}
  I.~L.~Buchbinder, S.~J.~Gates, Jr., H.~-S.~Goh, W.~D.~Linch, III, M.~A.~Luty, S.~-P.~Ng, J.~Phillips,
  %``Supergravity loop contributions to brane world supersymmetry breaking,''
  Phys.\ Rev.\  {\bf D70 } (2004)  025008.
  [hep-th/0305169].
  
%\bibitem{Gates:1983nr}
%  S.~J.~Gates, M.~T.~Grisaru, M.~Rocek, W.~Siegel,
%  %``Superspace Or One Thousand and One Lessons in Supersymmetry,''
%  Front.\ Phys.\  {\bf 58 } (1983)  1-548.
%  [hep-th/0108200].
  
\bibitem{Wess:1992cp}
  J.~Wess, J.~Bagger,
  ``Supersymmetry and supergravity,''
  Princeton, USA: Univ. Pr. (1992) 259 p.
  
\bibitem{Paccetti:2004ri}
  F.~Paccetti Correia, M.~G.~Schmidt, Z.~Tavartkiladze, ,
  %``Superfield approach to 5D conformal SUGRA and the radion,''
  Nucl.\ Phys.\  {\bf B709 } (2005)  141-170.
  [hep-th/0408138].

\bibitem{Abe:2004ar}
  H.~Abe, Y.~Sakamura,
  %``Superfield description of 5-D supergravity on general warped geometry,''
  JHEP {\bf 0410 } (2004)  013.
  [hep-th/0408224].

\end{thebibliography}
\end{document}